\documentclass[a4paper,twocolumn,11pt]{quantumarticle}
\pdfoutput=1
\usepackage[utf8]{inputenc}
\usepackage[english]{babel}
\usepackage[T1]{fontenc}
\usepackage{amsmath}
\usepackage{amsfonts}
\usepackage{amssymb}
\usepackage[numbers]{natbib}
\usepackage{bbold}
\usepackage{indentfirst}
\usepackage{mathtools}

\usepackage{tikz}
\usepackage{lipsum}

\usepackage{hyperref}

\newcommand{\ket}[1]{| #1 \rangle}
\newcommand{\bra}[1]{\langle #1 |}

\DeclareMathOperator{\Tr}{Tr}

\begin{document}

\title{Towards objectivity of classical reference frames in quantum mechanics}

\author{Rafał Ćwiek}
\email{rafal.cwiek@fuw.edu.pl}
\affiliation{University of Warsaw, Faculty of Physics, ul. Pasteura 5, 02-093 Warsaw, Poland}
\orcid{0000-0001-7436-8180}

\author{Jarosław K. Korbicz}
\email{jkorbicz@cft.edu.pl}
\orcid{0000-0003-2084-7906}
\affiliation{Center for Theoretical Physics, Polish Academy of Sciences, Aleja Lotników 32/46, 02-668 Warsaw, Poland}

\maketitle

\begin{abstract}
   Recent advances in our understanding of foundations of quantum mechanics have shown that information can be made objective through quantum states. Such objectification processes, predicted e.g. in a variety of quantum open systems, must accompany any realistic quantum-to-classical transition mechanism in order to reproduce the objective character of the classical limit of our world. However, so far only examples of simple, unstructured information, such as a value of an observable, have been studied. In this work we show that a more complicated form of information, given by a Cartesian reference frame, can also be made (at least partially) objective in quantum mechanics. The non-trivial internal gauge structure of reference frames, given by the transformation group, leads to a more general form of objectivity, where all observers see the same but modulo their relative orientations, like it happens in modern understanding of geometry. This opens a way to extend quantum objectivity beyond simple scenarios and possibly link it to the foundations of modern geometry.             
\end{abstract}

The possibility that the perceived objectivity of the classical world is in fact a specific property of quantum states has been an intriguing idea in modern physics, challenging traditional notions of reality and perception; see e.g. \cite{Zurek2009, Horodecki2015, LePhysRevLett2019, Korbicz2021, Zurek2025} for an introduction to different approaches, such as quantum Darwinism, Spectrum Broadcast Structures, and Strong Darwinism. Central to those approaches are the notions of information and objectivity, formulated using precise terms of modern quantum information science. This allows to successfully explain the objective part of the quantum-to-classical transition in the language of information proliferation and extraction \cite{Zurek2009}. Information studied so far has been 
of a simple structure such as a value of some observable, e.g. position \cite{PhysRevLett.101.240405, paz2009, Korbicz2014, GenRelativGravit2017.49:152, PhysRevA.109.052204} or spin \cite{PhysRevA.72.052113, lampo2017, mironowicz2018, PhysRevLett.123.140402,  Entropy2021_23(8)_995, PhysRevLett.128.010401}.

However, not all relevant information is of this simple structure. For example, reference frames play an important role in quantum mechanics \cite{bartlett2007, quantumframes}. In particular, the orientation in space or a $3D$ Cartesian reference frame. Quantum communication of the latter has been developed in \cite{bagan, Chiribella_2004}. A reference frame, represented by a $SO(3)$ rotation group element, is encoded in a state of $N$ spins $1/2$ (qubits). One party, Alice, prepares her system in a specifically tailored state and sends it to the other party, Bob. He then uses generalized quantum measurements (POVM - Positive Operator-Valued Measure) to decode the  group element that will allow him to align his frame to the otherwise unknown frame of Alice. By optimizing encoding/decoding, the average error can be shown to scale like $1/N^2$ \cite{Chiribella_2004}. 

In this work, we generalize this scheme by embedding it into an objectivity scenario. The goal is to not only transmit the reference frame to a single recipient but to make it available to multiple observers.  For that purpose we will use Spectrum Broadcast Structures (SBS)
\cite{Korbicz2014, Horodecki2015, Korbicz2021}, which is a state-oriented approach to objectivity, stronger than the original quantum Darwinism \cite{Horodecki2015,LePhysRevLett2019}. SBS are quantum state structures encoding a certain operational notion of objectivity of a state of the system of interest. At its heart lie two intuitive assumptions \cite{Zurek2009}:
\begin{enumerate}
    \item \emph{Agreement}  All observers see the same state. 
    \item \emph{Non-disturbance} The observations  do not disturb neither the observed system nor other observers.
\end{enumerate}
With some additional clarifications, most notably concerning the precise meaning of non-disturbance, a unique, idealized  state structure compatible with the above assumptions can be derived \cite{Horodecki2015, Korbicz2021}:
\begin{equation} \label{sbs}
    \rho_{SBS}=\sum_m p(m) \ket m \bra m \otimes \bigotimes_{i=1}^k \rho_{m}^i
\end{equation}
and
\begin{equation}\label{ort}
    \rho_m^{i} \perp \rho_{m'}^i\ \  \text{for every $i$ and $m\ne m'$.} 
\end{equation}
Here, $m$ labels a property that we want to make objective, encoded in the eigenstates $\ket m$ of the corresponding observable, $\rho_{m}^i$ are local states seen by the $i$th observer, and $p(m)$ is a probability distribution, which takes into account that in general physical properties are not sharply defined. Thanks to condition \eqref{ort}, i.e. $\rho_{m}^i$ have orthogonal supports for different $m$ and hence are perfectly disntinguishable, the structure \eqref{sbs} fulfills both objectivity conditions, with the measurements defined by the projections onto the orthogonal supports of $\rho_{m}^i$. In particular, the probability that  observers measure simultaneously values $m_1,\dots,m_k$ is
\begin{align}\label{zgoda}
    p(m_1,\dots,m_k)=\sum_m p(m) \delta_{mm_1}\cdots \delta_{mm_k},
\end{align}
which is the mathematical expression of the agreement condition: All the observers measure the same value of \(m\); otherwise, the probability is zero, and thus they agree on which state $\ket m$ the system is in. Loosely speaking, SBS states are sorts of reference frames for information being made objective.

In what follows, we will put the frame transmission scheme into the SBS structure \eqref{sbs}, using some methods of general quantum reference frames \cite{bartlett2007}. This leads to a novel aspect, which we call "covariant agreement". It reflects the nature of the reference frame information, where there is a symmetry/gauge group transforming one frame to another, and which is quite different from the "absolute" character of simple information like the value $m$ of some observable above. In case of the latter, all observers are supposed to agree on the value $m$, while the former allows a gauge transformation between the values obtained by different observers. We also analyze quantitatively the non-disturbance condition and show that actually it is not satisfied -- the post measurement state is significantly disturbed in the studied scenario. Despite this failure, our work sets a foot on the new ground of objectivity of geometric information. As a by-product, the disturbance analysis quantifies the disturbance of the original protocols \cite{bagan, Chiribella_2004}. 

The work is organized as follows. In Section \ref{sec:singleobserver} we briefly review the transmission scheme of \cite{bagan, Chiribella_2004}, supplementing it with explicit derivations of several quantities that will be needed later. Section \ref{sec:manyobservers} is the main part of the work, where we introduce the SBS-like construction for reference frames and the covariant agreement. In Section \ref{sec:agreement} we show that in the limit of large resource $N\to \infty$, the covariant agreement is indeed reached. In Section \ref{sec:disturbance} we analyze the asymptotic disturbance. We conclude with some additional remarks in Section \ref{sec:summary}. As derivations of our results are quite long and technical, they are presented in full detail in the Appendices.

\section{Transmission of a Reference Frame to a Single Observer}
\label{sec:singleobserver}
We will use the frame communication protocol developed in \cite{bagan, Chiribella_2004}. We review it here briefly for completeness, following \cite{Chiribella_2004} and supplementing with additional calculations, needed later for our construction. The protocol is the following. Alice aims to communicate an orthogonal trihedron, i.e. three mutually orthogonal unit vectors constituting a 3$D$ reference frame, to Bob using a quantum states. Specifically, she prepares a system of \(N\) elementary spins $1/2$ in a quantum state \( | A \rangle \), which is related to her orthogonal trihedron 
\begin{align}
    \mathbf{n}^{A} = \left( \hat{e}_x^A, \hat{e}_y^A, \hat{e}_z^A \right) \, .
\end{align}
Then she transmits the state \( | A \rangle \) to Bob, whose orthogonal trihedron $\mathbf{n}^{B} = \left( \hat{e}_x^B, \hat{e}_y^B, \hat{e}_z^B \right)$ is related to Alice's via  some unknown \(\mathrm{SO(3)} \) rotation 
\begin{align}
    \mathbf{n}^{A} = R_0 \mathbf{n}^{B} \, .
\end{align}
Rather than working with $SO(3)$ rotations, it is easier to work with the corresponding $SU(2)$ transformations: \( R_0 = \mathcal{R}(g_0) \), where \(g_0\) is any of the two corresponding elements of the \(\mathrm{SU(2)}\) group, which differ only in sign, i.e. \( \mathcal{R}(g_1) = \mathcal{R}(g_2) \iff g_1 = \pm g_2 \). The well-known properties of the mapping \( \mathcal{R}: \mathrm{SU(2)} \to \mathrm{SO(3)} \) are recalled in detail for the reader's convenience in Appendix \ref{appendix:su2group}. With respect to Bob's orthogonal trihedron, Alice's reference state \( | A \rangle \) will be seen as a rotated state 
\begin{align}\label{02jdn3n3idn3di9m2sz}
    | A(g_0) \rangle = U(g_0) | A \rangle \, .
\end{align}
By performing a measurement on the state \( | A(g_0) \rangle \), Bob will try to find out how his reference frame is related to Alice's one, i.e. he will try to estimate \( R_0 \) with as low error as possible. We stress that the communication scheme does not require any shared or pre-established reference frame. Below we summarize the main steps, while  a detailed overview can be found in Appendix \ref{appendix:chiribella}.

We recall that the Hilbert space of $N$ spins $1/2$, \( \mathcal{H}_N \), can be decomposed into a sum of $SU(2)$-invariant, irreducible subspaces:
\begin{align}\label{HN}
    \mathcal{H}_N =  \mathcal{H}_{\frac{1}{2}}^{\otimes N} = \bigoplus_{j=j_0}^{J} \bigoplus_{\alpha=1}^{n_j} \mathcal{H}_{j \alpha} \, ,
\end{align}
where \( j \) is the spin, \( j_0 = 0\) for \(N\) even and \( j_0 = \frac{1}{2}\) for \(N\) odd, \( J = \frac{N}{2}\) is the maximum spin, \( \alpha \) labels different equivalent representations with the same \( j \), and \( n_j \) is their multiplicity (see Eq. \eqref{nj}). In each \( \mathcal{H}_{j \alpha} \), \(\dim \mathcal{H}_{j \alpha}=2j+1\), there is the usual orthonormal basis \( |j\alpha, m\rangle \) of the eigenvectors of  \(J_z \in \mathfrak{su}(2) \). Following \cite{Chiribella_2004}, the Alice's reference state \( | A \rangle \in \mathcal{H}_N\) is chosen in the following form in order to maximize the transmission fidelity:
\begin{align}\label{stsrgu6ygr67i3446j}
    | A \rangle = \sum_{j=j_0}^{J-1} \sum_{\alpha=1}^{2j+1} \frac{A_j}{\sqrt{2j+1}} | j \alpha, m(\alpha) \rangle \, ,
\end{align}
where \( m(\alpha) \) is some injective function of \( \alpha \). We have arbitrarily chosen the component of \( | A \rangle \) in \( \mathcal{H}_{J 1} \) as equal to zero, \( A_J=0 \), because in the limit \(N \to \infty\), which is of interest here, the contribution of this representation can be neglected \cite{Chiribella_2004}. Bob wants to extract the rotation \(R_0\), or equivalently the $SU(2)$ element $g_0$, from the state \( | A (g_0) \rangle \). For that he employs a generalized (POVM) measurement \( \{ M(g') \}_{g' \in \mathrm{SU(2)}} \), i.e. a set of positive operators on \( \mathcal{H}_N \) satisfying the normalization $\int_{\mathrm{SU(2)}} \mathrm{d}g' M(g') = \mathbb{1}$, where \( \mathrm{d}g' \) is the normalized Haar measure on \( \mathrm{SU(2)} \). The probability density of estimating \( g' \) when the measured state is \( | A (g_0) \rangle \) is given by the Born rule 
\begin{align}\label{pcond}
    p(g'|g_0) = \mathrm{Tr} \left[ M(g')|A(g_0) \rangle \langle A(g_0)| \right] \, .
\end{align}
The efficiency of the communication strategy is characterized by the transmission error
\begin{align}\label{egg0}
    e(g', g_0) = \sum_{\alpha \in  \{ x, y, z \}} \left| \mathcal{R}(g') \hat{e}^A_\alpha - \mathcal{R}(g_0) \hat{e}^A_\alpha \right|^2
\end{align}
and its average value
\begin{align}\label{w983hd32m923dk20dA}
    \langle e \rangle = \int_{\mathrm{SU(2)}} \mathrm{d}g_0 \int_{\mathrm{SU(2)}} \mathrm{d}g' p(g' | g_0) e(g', g_0) \, .
\end{align}
The goal is now to find both the encoding state $|A\rangle$ and the POVM $M(g')$, minimizing  \eqref{w983hd32m923dk20dA}. First, the POVM is found using the maximum likelihood strategy, i.e. maximizing the peak \( p(g_0 | g_0) \) in the distribution \( p(g' | g_0) \). Because of the  $SU(2)$-invariance of \eqref{egg0} \cite{Holevo1982}, it is enough to consider covariant POVM's, i.e. $M(g') = U(g') \Xi U^{\dagger}(g')$, where \( \Xi \) is a positive operator. Combining the above with the fact that Alice’s reference vector \eqref{stsrgu6ygr67i3446j} lies in the following invariant subspace $\mathcal{K} = \bigoplus_{j=j_0}^{J-1} \bigoplus_{\alpha = 1}^{2j+1} \mathcal{H}_{j \alpha}$ of $\mathcal{H}_N$, leads to a simple form of the optimal POVM on \( \mathcal{K} \) \cite{Chiribella_2004_inny, Chiribella_2004}: $\Xi = | B \rangle \langle B|$, where
\begin{align}\label{mdj7suhe56wyegds2}
    | B \rangle = \sum_{j=j_0}^{J-1} \sum_{\alpha=1}^{2j+1} \sqrt{2j+1} \big| \, j \alpha, m(\alpha) \rangle \, 
\end{align}
and as a result \eqref{pcond} reads 
\begin{align}\label{pcondfromscalarproduct}
    p(g' | g_0) = \left| \langle A(g_0) | B(g') \rangle \right|^2 \, ,
\end{align}
where
\begin{align}\label{mx39xnj39v548z20fafU}
    | B(g') \rangle = U(g') | B \rangle \, .
\end{align}
What is left is the optimization of the coefficients \( \{ A_j \}\) in \eqref{stsrgu6ygr67i3446j}. It was shown in \cite{Chiribella_2004} using the Clebsch-Gordan theory that the minimum average error \( \langle e \rangle \) corresponds to the largest eigenvalue of the following tridiagonal matrix:
\begin{align}
    M =
    \begin{bmatrix}
    \zeta & 1 & 0 & 0 & \dots  & 0 \\
    1 & 1 & 1 & 0 & \dots  & 0 \\
    0 & 1 & 1 & 1 & \dots  & 0 \\
    0 & 0 & 1 & 1 & \dots  & 0 \\
    \vdots & \vdots & \vdots & \vdots & \ddots & \vdots \\
    0 & 0 & 0 & 0 & \dots  & 1
    \end{bmatrix} \, ,
\end{align}
where we omitted the row and column corresponding to  \( \mathcal{H}_{J 1} \) due to our assumption that \( A_J=0 \). Above, \( \zeta = 0(1) \) for even (odd) values of \( N \). Estimating the largest eigenvalue of \( M\), finally gives the desired error estimate \cite{Chiribella_2004}:
\begin{align}
    \langle e \rangle \sim \frac{8\pi ^2}{N^2}.
\end{align}

In what follows, we generalize the above transmission scheme to multiple receivers. The crucial role will be played by the conditional probability \eqref{pcondfromscalarproduct}, which was not explicitly calculated in the previous works, as only estimates were sufficient. We calculate it below. First, we explicitly determine the normalized largest eigenvalue eigenvector \( \mathbf A \) of \(M\) which will give the optimal \( \{ A_j \} \), since \cite{Chiribella_2004}
\begin{align}\label{wektorAT}
    \mathbf A =
    \begin{bmatrix}
    A_{j_0} & A_{j_0+1} & \ldots & A_{J-1} \\
    \end{bmatrix} \, .
\end{align}
The size of the matrix \( M \) is \( n \times n\), where
\begin{align}\label{ericn39cn39cn33}
    n = J - j_0 =
    \begin{cases} 
    J - \frac{1}{2}    & \mathrm{if} \; 2 \nmid N \\
    J                  & \mathrm{if} \; 2 \mid N 
    \end{cases} \, .
\end{align}
We make an approximation \( \zeta = 1 \) for both even and odd \( N \), justified for large $N$. Then \( M \) becomes a tridiagonal matrix, denoted \(M'\), with only ones on the diagonals for all values of \(N\). We show in Appendix \ref{appendix:ttm} that for such a matrix the normalized eigenvector corresponding to the largest eigenvalue has the following components (cf. \eqref{w0dn3dn2m0cm94b3bcn3}):
\begin{align}\label{njsuehdnbhhg2938uwj}
    v^l = \sqrt{\frac{2}{n+1}} \sin \left( \frac{l\pi}{n+1} \right) \, .
\end{align}
Relabeling its components to comply with \eqref{wektorAT}, where the first component is labeled $j_0$ rather than 1, we obtain:
\begin{align}\label{9phOVCFGHa2gv}
    A_j = v^{j+\frac{1}{2}} = \sqrt{\frac{2}{n+1}} \sin \left( \frac{j+\frac{1}{2}}{n+1} \pi \right)
\end{align}
for odd \(N\), and
\begin{align}\label{nwio4h93nc3a}
    A_j = v^{j+1} = \sqrt{\frac{2}{n+1}} \sin \left( \frac{j+1}{n+1} \pi \right)
\end{align}
for even \(N\), where \(n\) is given by \eqref{ericn39cn39cn33}. We remind the reader that while \eqref{9phOVCFGHa2gv} is an exact result, \eqref{nwio4h93nc3a} is only an approximation, since \(M'\) is only an approximation to \(M\) for large \(N\).

Subsequently, we calculate the scalar product \( \langle A \left( g \right) | B \left( g' \right) \rangle \). The detailed and lengthy calculations are presented in Appendix \ref{appendix:scalarproduct}; here, we briefly summarize the main steps. First, we note that
\begin{align}
    \langle A(g) | B(g') \rangle = \sum_{j=j_0}^{J-1} A_j \sum_{m=-j}^{j} D^j_{m, m}(g^{-1}g') \, ,
\end{align}
where \( D^j_{m, m} \left( g^{-1}g' \right) \) are matrix elements of the spin-$j$ representation (cf. \eqref{appfre9tgtg)HnjJmm}). The crucial observation here is that the scalar product depends on \(g\) and \(g'\) only through the characters of the representations, which leads to an expression
\begin{align}\label{dvnn93jmcp2/qjAw}
    \langle A(g) | B(g') \rangle = \sum_{j=j_0}^{J-1} A_j \frac{\sin \left[ (2j+1)\theta \right]}{\sin \theta} \, ,
\end{align}
which depends only on the relative angle \(\theta \in [0,\pi]\), given by
\begin{align}
    \cos\theta = \frac{1}{2} \Tr \left( g^{-1}g' \right) \, .
\end{align}
More generally, for an arbitrary \(g \in \mathrm{SU(2)} \), the corresponding hyperspherical angle \( \theta_g \in [0,\pi] \), cf. Appendix \ref{appendix:su2group},  is given by $\cos \theta_g = \frac{1}{2} \Tr \left( g \right)$. It is easy to see that \( \theta_{-g} = \pi - \theta_g \), and both of these angles correspond to the same \(SO(3)\)-rotation, i.e. \( \mathcal{R}(g)=\mathcal{R}(-g) \). Moreover,  \( \theta_g = 0 \) if and only if \( g = \mathbb{1} \). In particular, \( \theta = 0 \) in  \eqref{dvnn93jmcp2/qjAw} corresponds to \( g=g' \), and \( \theta = \pi \) corresponds to \( g=-g' \); both cases correspond to successful decoding of the transmitted reference frame. In both of these cases,  \( \mathcal{R}(g^{-1}g') = \mathbb{1} \), i.e. $\mathcal{R}(g) = \mathcal R(g')$. Equation \eqref{dvnn93jmcp2/qjAw} can be further simplified as follows (cf. \eqref{7uhgtghhgytyu67ghgmWd}):
\begin{align}\label{imunycvtnxw}  
\begin{split}
    \langle A(g) | B(g') \rangle &= \sqrt{\frac{2}{n+1}} \frac{\sin \left[ (J+j_0) \theta \right]}{\sin \theta} \\
    &\times \dfrac{\sin \left( \dfrac{\pi}{n+1} \right) \cos \left[ (n+1) \theta \right]}{\cos 2 \theta - \cos \left( \dfrac{\pi}{n+1} \right)} \, ,
\end{split}
\end{align}
where, as before, \(n=J-j_0\). Finally, from \eqref{imunycvtnxw} and \eqref{pcondfromscalarproduct}, we obtain the desired expression for the probability density:
\begin{align}\label{pgg'}
\begin{split}
    p(g' | g) &= \frac{2}{n+1} \frac{\sin^2 \left[ (J+j_0) \theta \right]}{\sin^2 \theta} \\
    &\times \dfrac{\sin^2 \left( \dfrac{\pi}{n+1} \right) \cos^2 \left[ (n+1) \theta \right]}{\left[ \cos  2 \theta - \cos \left( \dfrac{\pi}{n+1} \right) \right]^2} \, .
\end{split}
\end{align}
We emphasize that this is an exact result, given the definitions of vectors \(|A\rangle\) and \(|B\rangle\) from \eqref{stsrgu6ygr67i3446j}, \eqref{mdj7suhe56wyegds2}, \eqref{9phOVCFGHa2gv}, and \eqref{nwio4h93nc3a}.

\section{Transmission of a Reference Frame to Many Observers}
\label{sec:manyobservers}
Our goal is to objectify Alice's reference frame and communicate it to \(k\) Bobs, playing the role of the observers as per the basic objectivity setup \cite{Zurek2009, Korbicz2021}. The idea is to take the communication protocol summarized in the previous section and put it into an SBS-type structure \eqref{sbs}. First of all, we can identify Alice's reference frame with an element of the \(\mathrm{SO(3)}\) group by the usual construction. To do so, we introduce a fixed orthogonal trihedron \( \mathbf{n} = \left( \hat{e}_x, \hat{e}_y, \hat{e}_z \right) \). Then Alice's orthogonal trihedron is given by 
\begin{align}\label{ncdjiden3drf34}
    \mathbf{n}^{A} = \left( \hat{e}_x^A, \hat{e}_y^A, \hat{e}_z^A \right) = R_A \mathbf{n} \, .
\end{align}
Each of the Bobs initially holds a reference frame $\mathbf{n}^{i} = \left( \hat{e}_x^i, \hat{e}_y^i, \hat{e}_z^i \right)$, which they want to align to Alice's one through a rotation
\begin{align}\label{cme89m8x3mc49}
    \mathbf{n}^{A} = R_A R_i \mathbf{n}^{i} \, ,
\end{align}
where \( i \in \overline{1,k} \) and \( R_i \in \mathrm{SO(3)} \) are fixed by $\mathbf{n}^{i}$. Alice constructs the state \( | A \rangle \) with respect to her orthogonal trihedron \( \mathbf{n}^{A} \) and sends it to each Bob. With respect to the \(i\)th Bob's orthogonal trihedron \( \mathbf{n}^{i} \), it will be seen as a rotated state 
\begin{align}\label{Agg}
    | A (g_A g_i) \rangle = U(g_Ag_i)|A\rangle \, ,  
\end{align}
where \(g_i\) is any of the two elements of the \(\mathrm{SU(2)}\) group satisfying \( R_i = \mathcal{R}(g_i) \). Then each Bob will try to find out how his orthogonal trihedron is related to Alice's one, i.e., he will try to estimate \( R_A R_i \). Let us emphasize that, just like for one observer, such a communication scheme works without a need of any common reference frame. In particular, the orthogonal trihedron \( \mathbf{n} \) does not have to be known to any of the Bobs. The objectivity of Alice's reference frame, identified with \( R_A \in \mathrm{SO(3)}\), can be understood as follows: If the \(i\)th Bob claims that Alice's reference frame is described by \(R_A R_i\) with respect to his frame, while the \(j\)th Bob claims that Alice's reference frame is described by \(R_A R_j\), then they can compare the two results, obtaining
\begin{align}\label{RR}
    (R_A R_j)^{-1} R_A R_i = R_j^{-1} R_i \, ,
\end{align}
which corresponds exactly to their relative orientation: $\mathbf{n}^{j} = R_j^{-1} R_i \mathbf{n}^{i}$, which follows from \eqref{cme89m8x3mc49}. Thus, they conclude that they agree on Alice's reference frame up to their relative rotation. This is a new form of a "covariant" agreement as opposed to the "absolute" one considered so far in the objectivity studies \cite{Zurek2009, Horodecki2015, Korbicz2021}. There the demand was that each observer measures exactly, in the idealized picture, the same value of the observable that is supposed to be objective.

Instead of assuming a precisely defined frame $g_A$, we will allow for some indefiniteness, described by a probability distribution $p_0$ over all possible frames. We need to introduce a Hilbert space \( \mathcal{H}_R \), in which all possible Alice's reference frames will be faithfully encoded. For that, we will use an idea from the general theory of reference frames in quantum mechanics \cite{bartlett2007} of how to include a reference frame in a general quantum description.  This is done using the left regular representation \cite{leftregular} of the corresponding group, acting in \(\mathcal{H}_R=L^2(G) \). We introduce states \( | g \rangle \), labeled by all \( g \in \mathrm{SU(2)} \). We define the \(\mathrm{SU(2)}\) action \(U_R\) via the left multiplication
\begin{align}
    U_R (h) | g \rangle = | hg \rangle,
\end{align}
which corresponds to a frame change and ensures that there are no non-trivial invariant subspaces. To further guarantee that this quantum system encodes perfectly all possible reference frames, the states \( | g \rangle \) must be fully distinguishable, i.e. orthogonal for different group elements:
\begin{align}\label{diracdeltanormalization}
    \langle g | h \rangle = \delta(g,h) \, ,
\end{align}
where \( \delta (g,h) \) is the Dirac delta function on the \( \mathrm{SU(2)} \) group, defined by the condition that
\begin{align}\label{delta}
    \int\displaylimits_{\mathrm{SU(2)}} \mathrm{d}g \, \delta (g,h) \varphi(g) = \varphi(h)
\end{align}
for every \( \varphi \in C^{\infty}(\mathrm{SU(2)}) \). These properties define the left regular representation of the \( \mathrm{SU(2)} \) group \cite{bartlett2007, leftregular}. Of course, the states are not normalizable due to the condition \eqref{delta}, in the similar way as the eigenstates of a continuous observable are not. We will not address this well known problem here and continue working with the states $|g\rangle$ to convey the basic physical idea behind our proposal. We thus consider the following Hilbert space:
\begin{align}
    \mathcal{H}= \mathcal{H}_R \otimes \bigotimes_{i=1}^{k} \mathcal{H}_N \, ,
\end{align}
where \( \mathcal{H}_N \) is the Hilbert space of $N$ spins $1/2$, cf. \eqref{HN}, held by each Bob. Motivated by the SBS state \eqref{sbs}, we then construct the following states on \( \mathcal{H} \) that will serve to objectify Alice's reference frame:
\begin{align}\label{brtfgh7uyg687hjhTGhj7}
\begin{split}
    \rho = \int\displaylimits_{\mathrm{SU(2)}} \mathrm{d}g \, p_0 &\left[ \mathcal{R}(g) \right] | g \rangle \langle g | \otimes\\
    &\otimes \bigotimes_{i=1}^{k} | A(g g_i) \rangle \langle A(g g_i) | =\\
    = \int\displaylimits_{\mathrm{SU(2)}} \mathrm{d}g \, p_0 &\left[ \mathcal{R}(g) \right] | g \rangle \langle g | \otimes\\
    \otimes \bigotimes_{i=1}^{k} \, & U(g) | A(g_i) \rangle \langle A(g_i) | U(g)^\dagger \, ,
\end{split}
\end{align}
where $p_0\left[ \mathcal{R}(g) \right]$ is the probability distribution characterizing indefiniteness in Alice's reference frame definition, satisfying
\begin{align}
    \int\displaylimits_{\mathrm{SO(3)}} \mathrm{d}R \, p_0(R) = \int\displaylimits_{\mathrm{SU(2)}} \mathrm{d}g \, p_0 \left[ \mathcal{R}(g) \right] = 1 \, ,
\end{align}
(the first equality is in fact satisfied by all measurable functions on the \( \mathrm{SO(3)} \); cf. \eqref{yvgjbh547687bcfgVGHJNH7d}). Interestingly, modulo the probability distribution $p_0$, the state \eqref{brtfgh7uyg687hjhTGhj7} looks like a  frame-invariant version of the state $\bigotimes_{i}| A(g_i) \rangle \langle A(g_i) |$, according to the frame quantization procedure of \cite{bartlett2007}:
\begin{align}\label{rhoinv}
    \rho^{inv} = \int_G \mathrm d g | g \rangle \langle g | \otimes U(g) \rho U(g)^\dagger
\end{align}
(in the recent language this nomenclature can be somewhat misleading as it is different from more straightforward, albeit also more limited, frame quantization procedures of e.g. \cite{quantumframes}). However, we reached \eqref{brtfgh7uyg687hjhTGhj7} from a different direction, i.e. constructing an SBS-like state and using transformation law \eqref{Agg} that appears naturally in our scenario. The left regular representation space was used just as a tool to encode all possible frames of Alice. More importantly, the presence of $p_0$ implies our state \eqref{brtfgh7uyg687hjhTGhj7} need not be frame invariant, which was the main reason for introducing \eqref{rhoinv}. Nevertheless, there seems to be some interplay between the frame-objectivization and frame internalization.

Bobs are now carrying local measurements on their respective subsystems in order to extract the necessary group element and align their frames to that of Alice. The measurements are independent, as it is demanded by the objectivity definition \cite{Horodecki2015,Korbicz2021} and given by the POVM \eqref{mdj7suhe56wyegds2}, \eqref{mx39xnj39v548z20fafU} described in the previous Section. Thus, the composite measurement reads
\begin{align}\label{compositePOVM}
    M(g'_1, \ldots, g'_k) = \openone_R \otimes \bigotimes_{i=1}^{k} | B(g'_i) \rangle \langle B(g'_i) | \, . 
\end{align}
The joint probability that the \(i\)th Bob estimates \( g_i' \) when the measured state is \( \rho \) equals
\begin{align}\label{btngyjjrth5467yubhHGdfgh6}
\begin{split}
    &p(g_1', \dots, g_k') = \mathrm{Tr} \left[ M(g'_1, \ldots, g'_k) \rho \right]=\\
    =& \int\displaylimits_{\mathrm{SU(2)}} \mathrm{d}g \, p_0 \left[ \mathcal{R}(g) \right] \mathrm{Tr} \left( |g \rangle \langle g| \right)  \prod_{i=1}^{k} p( g_i' | g g_i ) \, ,
\end{split}
\end{align}
where
\begin{align}\label{probdensity}
    p( g_i' | g g_i ) = |\langle A(g g_i) | B(g_i') \rangle |^2 \, 
\end{align}
and we calculated it in \eqref{pgg'}. As we mentioned, the state \( \rho \) introduced above is, in a sense, a heuristic state because it involves an infinite trace, making the probability density \( p(g_1', \dots, g_k') \) ill-defined. For the purpose of this work, we will bypass this issue and redefine the probability density as
\begin{align}\label{fbhjhgjtj457678ghghdfhjgYTp7}
    p(g_1', \dots, g_k') = \int\displaylimits_{\mathrm{SU(2)}} \mathrm{d}g \, p_0 \left[ \mathcal{R}(g) \right] \prod_{i=1}^{k} p( g_i' | g g_i ) \, .
\end{align}
We can rewrite the above probability density in terms of the \( \mathrm{SO(3)} \) group. For \( R,R' \in \mathrm{SO(3)} \), we define
\begin{align}\label{rrbty45677hygjbfsfF}
    p(R'|R) \equiv p(g'|g) \, ,
\end{align}
where \(g\) and \(g'\) are any elements of the \(\mathrm{SU(2)}\) group satisfying \( R=\mathcal{R}(g) \) and \( R'=\mathcal{R}(g') \). Such a probability density \(p(R'|R)\) is well-defined because \( \langle A(g) | B(g') \rangle \) can only differ in sign for different choices of \(g\) and \(g'\), always resulting in the same \(p(g'|g)\). Therefore, Eq. \eqref{fbhjhgjtj457678ghghdfhjgYTp7} can be rewritten as follows (cf. \eqref{yvgjbh547687bcfgVGHJNH7d}):
\begin{align}\label{bgh58768hjhnHNJJ78hyu4}
    p(R_1', \dots, R_k') = \int\displaylimits_{\mathrm{SO(3)}} \mathrm{d}R \, p_0(R) \prod_{i=1}^{k} p( R_i' | R R_i ) \, .
\end{align}

\section{Asymptotic Covariant Agreement of Multiple Observers}
\label{sec:agreement}
We want to show that, in the limit of a large number of spins \( N \to \infty \), or equivalently $J\to \infty$, the reference frames determined by different observers from \eqref{brtfgh7uyg687hjhTGhj7} will asymptotically agree with each other, modulo the relative orientation \eqref{RR}. First, we prove that if \( f \in C^{\infty} \left[ \mathrm{\mathrm{SO(3)}} \right] \) is a test function and \( g' \) is a fixed element of the \( \mathrm{SU(2)} \) group, then the following holds:
\begin{align}\label{nyguj87uitnghjhfHGHJjnh7yu78}
    \lim_{J \to +\infty} \int\displaylimits_{\mathrm{SU(2)}} \mathrm{d}g \, p( g' | g ) f \left[ \mathcal{R} (g) \right] = f \left[ \mathcal{R} \left( g' \right) \right] \, ,
\end{align}
where $p( g' | g )$ is given by \eqref{pgg'}. A detailed derivation of this fact can be found in Appendix \ref{appendix:delta}, see Eq. \eqref{tdvryhgtfvhyugvfy5uyu6576yG}. Using identification \eqref{yvgjbh547687bcfgVGHJNH7d}, we can rewrite \eqref{nyguj87uitnghjhfHGHJjnh7yu78} on $SO(3)$
\begin{align}
    \lim_{J \to +\infty} \int\displaylimits_{\mathrm{\mathrm{SO(3)}}} \mathrm{d}R \, p( R' | R ) f(R) = f(R') \, ,
\end{align}
which can be restated in terms of the Dirac delta function as
\begin{align}\label{deltaRR'}
    \lim_{J \to +\infty} p( R' | R ) = \delta ( R', R ) \, ,
\end{align}
i.e. in the limit \( J \to \infty \), Alice’s reference frame can be transmitted to Bob without an error. Having established \eqref{deltaRR'}, we  analyze the large-spin limit of  \eqref{bgh58768hjhnHNJJ78hyu4}. Let us take a test function from the total group space, i.e. from the $k$ copies of $SO(3)$, describing $k$ observers
\(    f_{tot} \in C^{\infty} \left[ \bigtimes_{i=1}^{k} \mathrm{\mathrm{SO(3)}} \right]\).
Using \eqref{deltaRR'} and the fact that product functions are dense in the space of all $f_{tot}$, we prove one of our central results (see Eq. \eqref{retygnjk6778Gyuhj6ASc} of Appendix \ref{appendix:delta}):
\begin{align}
\begin{split}
    \lim_{J \to +\infty }\idotsint\displaylimits_{\mathrm{\mathrm{SO(3)}} \; \, \; \mathrm{\mathrm{SO(3)}}} \mathrm{d}R'_1 \ldots \mathrm{d}R'_k \, p(R'_1, \ldots, R'_k) \\
    \times f_{tot} \left( R'_1, \ldots, R'_k \right) =\\ =\int\displaylimits_{\mathrm{\mathrm{SO(3)}}} \mathrm{d}R \, p_0(R) f_{tot} \left( R R_1, \ldots, R R_k \right) \, ,
\end{split}
\end{align}
which in terms of the Dirac delta functions reads
\begin{align}
\begin{split}\label{agreement}
    \lim_{J \to +\infty } p&(R'_1, \ldots, R'_k) =\\
    &= \int\displaylimits_{\mathrm{\mathrm{SO(3)}}} \mathrm{d}R \, p_0(R) \prod_{i=1}^{k} \delta (R R_i, R_i') \, .
\end{split}
\end{align}
This is the expression of the covariant agreement, including a classical  indefiniteness (noise) in the transmitted frame, cf. \eqref{zgoda}. Indeed, if Alice's reference frame is described by \( R \in \mathrm{SO(3)} \), then according to \eqref{agreement}, the \(i\)th Bob will obtain through the measurement the group element \( R R_i \in \mathrm{SO(3)}\), distributed with the same probability $p_0(R)$ as Alice's frame. But this is exactly the rotation needed to align his frame with Alice's one as per \eqref{cme89m8x3mc49}: If he applies the obtained rotation \( R R_i \) to his reference frame, he will end up in the reference frame
\begin{align}\label{nA}
    \mathbf{n}^i_{rotated} &= R R_i \mathbf{n}^i = R R_i (R R_i)^{-1} \mathbf{n}^{A} = \mathbf{n}^{A} \, ,
\end{align}
which is just Alice's reference frame. Let us stress again the novel, covariant character of the agreement described here: Each observer obtains a different group element $RR_i$ but they all will align to the same reference frame via \eqref{nA}.

\section{Asymptotic Disturbance}
\label{sec:disturbance}
Having discussed the agreement part of the objectivity, we now proceed to the non-disturbance. To assess how much the reference frame extraction disturbs the state \eqref{brtfgh7uyg687hjhTGhj7}, particularly in the limit of large \(J\), we evaluate the trace distance between the post-measurement and the original (pre-measurement) states. Following the interpretation of "non-disturbance" in the SBS framework \cite{Korbicz2021}, we focus on the average post-measurement state, averaged over possible measurement results. Since the trace distance is difficult to compute directly, we estimate it using the state fidelity and the Fuchs–van de Graaf inequalities \cite{fuchs}. A detailed derivation of the post-measurement state and the corresponding bounds on the trace distance is provided in Appendix \ref{appendix:disturbance}, while here we summarize only the main steps of the calculation.

The composite measurement, corresponding to the observers trying to extract the reference frame, is given by the product POVM \( M(g'_1, \ldots, g'_k) \) (cf. \eqref{compositePOVM}), and the corresponding  measurement operators:
\begin{align}
    \sqrt{M(g'_1, \ldots, g'_k)} = \openone_R \otimes \bigotimes_{i=1}^{k} \frac{| B(g'_i) \rangle \langle B(g'_i) |}{\| B \|} \, ,
\end{align}
where the norm \(\|B\|\) appears due to the square root operation. Strictly speaking, taking the square root of the effects \(M(g'_1, \ldots, g'_k) \) introduces an ambiguity up to a unitary rotation. The original proposal \cite{bagan, Chiribella_2004} did not resolve this ambiguity, as doing so was unnecessary for its purposes. In our interpretation, the decoding algorithm consists simply of projections onto the $\ket B$ vectors, without the application of any additional unitary rotations. Such unitary rotations would need to satisfy a hierarchy of consistency conditions arising from the partial traces of \(M(g'_1, \ldots, g'_k) \). With this assumption, the average post-measurement state is (cf. \eqref{postmstateapp}):
\begin{align}\label{postmstate}
    \rho' = \int\displaylimits_{\mathrm{SU(2)}} \mathrm{d}g \, p_0 \left[ \mathcal{R}(g) \right] | g \rangle \langle g | \otimes \sigma \, ,
\end{align}
where
\begin{align}\label{maindefsigma}
    \sigma \equiv \bigotimes_{i=1}^{k} \int\displaylimits_{\; \; \mathrm{SU(2)}} \mathrm{d}g'_i \, p( g_i' | g g_i ) \frac{| B(g'_i) \rangle \langle B(g'_i) |}{\| B \|^2} \, .
\end{align}
Our goal is to compute the trace distance between the post-measurement state \( \rho' \) (Eq. \eqref{postmstate}) and the pre-measurement state \( \rho \) (Eq. \eqref{brtfgh7uyg687hjhTGhj7}). The trace distance is defined as follows:
\begin{align}
   \frac{1}{2} \|\rho - \rho'\|_{1} \equiv \frac{1}{2} \Tr \left[ \sqrt{(\rho-\rho')^\dagger (\rho-\rho')} \right] \, .
\end{align}
The non-disturbance means that the states \(\rho\) and \(\rho'\) are nearly indistinguishable: \( \rho \approx \rho' \), which corresponds to \( \|\rho - \rho'\|_{1} \approx 0 \). As we calculated in Appendix \ref{appendix:disturbance}, the trace norm of \( \rho - \rho' \) is (cf. \eqref{apptracenorm}):
\begin{align}
    \left\| \rho - \rho' \right\|_{1} = \int\displaylimits_{\mathrm{SU(2)}} \mathrm{d}g \, p_0 \left[ \mathcal{R}(g) \right] \Tr \left( |g \rangle \langle g| \right) \left\| C(g) \right\|_{1} ,
\end{align}
where
\begin{align}\label{maindefC(g)}
    C(g) \equiv | \psi \rangle \langle \psi | - \sigma
\end{align}
and
\begin{align}\label{maindefpsi}
    | \psi \rangle \equiv \bigotimes_{i=1}^{k} | A(g g_i) \rangle \, .
\end{align}
Similarly to the case of calculating the joint probability density \( p(g_1', \dots, g_k') \) (see Eqs. \eqref{btngyjjrth5467yubhHGdfgh6} and \eqref{fbhjhgjtj457678ghghdfhjgYTp7}), we omit the  problematic term \( \Tr \left( |g \rangle \langle g| \right) \) and redefine the trace norm of \( \rho - \rho' \) as follows:
\begin{align}
    \left\| \rho - \rho' \right\|_{1} = \int\displaylimits_{\mathrm{SU(2)}} \mathrm{d}g \, p_0 \left[ \mathcal{R}(g) \right] \left\| C(g) \right\|_{1} \, .
\end{align}
Subsequently, we estimate the trace norm of the operator \( C(g) \) by bounding it in terms of the fidelity, using the Fuchs–van de Graaf inequalities (see \cite{fuchs} and \cite{Nielsen_Chuang_2010}). For arbitrary density matrices \(\rho_1\) and \(\rho_2\), the state fidelity is defined as
\begin{align}
    F(\rho_1, \rho_2) \equiv \left( \Tr \sqrt{\sqrt{\rho_1} \rho_2 \sqrt{\rho_1}} \right)^2 \, .
\end{align}
The lower bound is given by:
\begin{align}\label{fidelitylower}
     1 - F \left( |\psi \rangle \langle \psi | , \sigma \right) \leq \frac{1}{2} \Big\| | \psi \rangle \langle \psi | - \sigma \Big\|_{1} \, .
\end{align}
It is tighter than \( 1 - \sqrt{F \left( |\psi \rangle \langle \psi | , \sigma \right)} \) in \cite{fuchs}, because one of our states is pure \cite{Nielsen_Chuang_2010}. The upper bound is as follows:
\begin{align}\label{fidelityupper}
    \frac{1}{2} \Big\| | \psi \rangle \langle \psi | - \sigma \Big\|_{1} \leq \sqrt{1 - F \left( |\psi \rangle \langle \psi | , \sigma \right)} \, .
\end{align}
We then show that (see Eqs. \eqref{lambdafinal} and \eqref{fidelitylimit} of Appendix \ref{appendix:disturbance}):
\begin{align}
    \lim_{J \to +\infty} F \left( |\psi \rangle \langle \psi | , \sigma \right) = \lambda^k \, ,
\end{align}
where
\begin{align}
    \lambda \equiv 12 \pi^3 \int\displaylimits_{-\infty}^{+\infty} \dfrac{ \sin^4 x}{ x^2 \left( x^2 - \pi^2 \right)^4} \, \mathrm{d} x \approx 0.236 \, .
\end{align}
Thus, in the limit \( J \to \infty \), the trace distance between the post-measurement state and the pre-measurement state is bounded by the following expressions (cf. \eqref{tracedistanceestimated2}):
\begin{align}\label{tracedistancemain}
     1 - \lambda^k \leq \lim_{J \to +\infty} \frac{1}{2} \left\| \rho - \rho' \right\|_{1} \leq \sqrt{1 - \lambda^k} \, .
\end{align}
What matters most is the lower bound, as it separates the trace distance from zero, i.e. from the ideal non-disturbance. We see that even for a single Bob (\(k=1\)), there is already a significant disturbance, and this disturbance increases rapidly with the number of Bobs. In fact, as \( k \to \infty \), the trace distance approaches \(1\), indicating that the pre-measurement and post-measurement states become nearly orthogonal. Consequently, the upper bound on the trace distance becomes trivial in this regime. Thus, although the objectivization scheme, represented by \eqref{brtfgh7uyg687hjhTGhj7} and \eqref{compositePOVM}, asymptotically leads to a covariant form of the agreement condition, it fails in satisfying the non-disturbance criterion. We note, however, that the induced disturbance is local -- it affects only the local state of each observer. In any case, due to the failure of the non-disturbance condition, we can speak only of partial objectivity here.

As a side remark, when \(k=1\), the above calculations provide the disturbance analysis for the original transmission scheme -- an aspect that was not examined in the original works \cite{bagan, Chiribella_2004}; cf. Appendix \ref{appendix:disturbance} for more details.

\section{Final Remarks}
\label{sec:summary}
We have shown how to extend the notion of objectivity in quantum mechanics, developed and actively studied in the recent years, to more complicated forms of information such as a reference frame. Using as an example $3D$ Cartesian reference frames and the known mechanism of their quantum transmission \cite{bagan, Chiribella_2004}, we have proposed a way towards making them objective, using Spectrum Broadcast Structures (SBS) quantum states. 

The non-trivial gauge structure of the transmitted information, given by the group action that changes reference frames, forces the notion of agreement, the cornerstone of objectivity, to be generalized to what we call a covariant agreement. The measurements of the observers are not identical but connected via a reference frame rotation. This is in fact very well known in the geometry. Since the proclamation of the Erlangen program by Felix Klein \cite{Klein}, geometric objects, e.g. vectors, are seen to exist objectively, i.e. independently of  a reference frame. Viewed from different frames, they appear different but those are just different views of the same object, connected by an appropriate group transformation. We speculate that our ideas, particularly the covariant agreement, may be relevant for making a connection between the foundations of modern geometry and quantum mechanics, potentially leading to what one might call a quantum geometry. This is especially relevant given our belief that our results can be directly generalized to arbitrary reference frames in quantum mechanics, using e.g. the transmission mechanism of \cite{bartlett2007}.  

Of course, there remains a drawback in our scheme -- the disturbance experienced during the observation. This is not entirely surprising, as the disturbance was not relevant for the original reference frame transmission schemes \cite{bagan,Chiribella_2004}, which were focused on the maximal information extraction. As a result, there is some degradation of objectivity upon observation, in the sense of the local disturbance induced by each observer. This is somewhat similar to the quantum reference frame degradation \cite{bartlett2007}.  Although one can argue that the agreement is more important/interesting than the non-disturbance, it would be interesting to find new schemes where the non-disturbance is also obeyed, at least in some form.

Another interesting question is if the SBS-like states \eqref{brtfgh7uyg687hjhTGhj7} can arise (approximately) in the course of some evolution \cite{Korbicz2017}. The natural environment for the usual SBS states are open systems, where they were theoretically predicted to appear as a result of decoherence in all major models \cite{Korbicz2014, tuziemski2015, Korbicz2017, lampo2017, mironowicz2018, PhysRevA.109.052204}. Following the same path here would require a central system to be described by the Hilbert space of the left regular representation of $SU(2)$. This would be a rather exotic physical system. But even if states \eqref{brtfgh7uyg687hjhTGhj7} do not appear naturally, they provide a rather non-trivial extension of the notion of objectivity in quantum mechanics and show that more complicated forms of information can be in principle made quantum-objective. This also shows universality of the SBS structure (cf. \cite{PhysRevResearch.3.033148}). 

Finally, it would be interesting to further explore the connection between the frame objectivization \eqref{brtfgh7uyg687hjhTGhj7} and the frame inclusion \eqref{rhoinv} of \cite{bartlett2007}. The crucial difference between the two from a mathematical point of view lies in the presence of the probability distribution $p_0$, which breaks the group-invariance of the state  \eqref{brtfgh7uyg687hjhTGhj7}. Indeed, non-trivial distributions cannot be left-invariant -- otherwise, they would simply be constant functions on the group. Consequently, the SBS state \eqref{brtfgh7uyg687hjhTGhj7} cannot be directly interpreted in terms of the frame-independence, which is central to the frame-inclusion mechanism. Nevertheless, we believe there is an interpretational link between the two.

\section*{Acknowledgements}
We acknowledge the financial support of the Polish National Science Foundation (NCN) through the grant Opus 2019/35/B/ST2/01896, JKK acknowledges the support of the Quant-Era grant  ``Quantum Coherence Activation By Open Systems and Environments'' QuCABOoSE 2023/05/Y/ST2/00139.

\bibliographystyle{quantum}
\bibliography{bibliography}

\begin{thebibliography}{10}

\bibitem{Zurek2009}
Wojciech~Hubert Zurek.
\newblock ``Quantum darwinism''.
\newblock \href{https://dx.doi.org/10.1038/nphys1202}{Nature Physics {\bf 5}, 181--188}~(2009).

\bibitem{Horodecki2015}
R.~Horodecki, J.~K. Korbicz, and P.~Horodecki.
\newblock ``Quantum origins of objectivity''.
\newblock \href{https://dx.doi.org/10.1103/PhysRevA.91.032122}{Phys. Rev. A {\bf 91}, 032122}~(2015).

\bibitem{LePhysRevLett2019}
Thao~P. Le and Alexandra Olaya-Castro.
\newblock ``Strong quantum darwinism and strong independence are equivalent to spectrum broadcast structure''.
\newblock \href{https://dx.doi.org/10.1103/PhysRevLett.122.010403}{Phys. Rev. Lett. {\bf 122}, 010403}~(2019).

\bibitem{Korbicz2021}
J.~K. Korbicz.
\newblock ``Roads to objectivity: {Q}uantum {D}arwinism, {S}pectrum {B}roadcast {S}tructures, and {S}trong quantum {D}arwinism – a review''.
\newblock \href{https://dx.doi.org/10.22331/q-2021-11-08-571}{{Quantum} {\bf 5}, 571}~(2021).

\bibitem{Zurek2025}
Wojciech~Hubert Zurek.
\newblock ``Decoherence and quantum darwinism: From quantum foundations to classical reality''.
\newblock \href{https://dx.doi.org/https://doi.org/10.1017/9781009552868}{Cambridge University Press}. ~(2025).

\bibitem{PhysRevLett.101.240405}
Robin Blume-Kohout and Wojciech~H. Zurek.
\newblock ``Quantum darwinism in quantum brownian motion''.
\newblock \href{https://dx.doi.org/10.1103/PhysRevLett.101.240405}{Phys. Rev. Lett. {\bf 101}, 240405}~(2008).

\bibitem{paz2009}
Juan~Pablo Paz and Augusto~J. Roncaglia.
\newblock ``Redundancy of classical and quantum correlations during decoherence''.
\newblock \href{https://dx.doi.org/10.1103/PhysRevA.80.042111}{Phys. Rev. A {\bf 80}, 042111}~(2009).

\bibitem{Korbicz2014}
J.~K. Korbicz, P.~Horodecki, and R.~Horodecki.
\newblock ``Objectivity in a noisy photonic environment through quantum state information broadcasting''.
\newblock \href{https://dx.doi.org/10.1103/PhysRevLett.112.120402}{Phys. Rev. Lett. {\bf 112}, 120402}~(2014).

\bibitem{GenRelativGravit2017.49:152}
J.~K. Korbicz and J.~Tuziemski.
\newblock ``Information transfer during the universal gravitational decoherence''.
\newblock \href{https://dx.doi.org/10.1007/s10714-017-2319-3}{General Relativity and Gravitation {\bf 49}, 152}~(2017).

\bibitem{PhysRevA.109.052204}
Tae-Hun Lee and Jaros\l{}aw~K. Korbicz.
\newblock ``Encoding position by spins: Objectivity in the boson-spin model''.
\newblock \href{https://dx.doi.org/10.1103/PhysRevA.109.052204}{Phys. Rev. A {\bf 109}, 052204}~(2024).

\bibitem{PhysRevA.72.052113}
F.~M. Cucchietti, J.~P. Paz, and W.~H. Zurek.
\newblock ``Decoherence from spin environments''.
\newblock \href{https://dx.doi.org/10.1103/PhysRevA.72.052113}{Phys. Rev. A {\bf 72}, 052113}~(2005).

\bibitem{lampo2017}
Aniello Lampo, Jan Tuziemski, Maciej Lewenstein, and Jaros\l{}aw~K. Korbicz.
\newblock ``Objectivity in the non-markovian spin-boson model''.
\newblock \href{https://dx.doi.org/10.1103/PhysRevA.96.012120}{Phys. Rev. A {\bf 96}, 012120}~(2017).

\bibitem{mironowicz2018}
P.~Mironowicz, P.~Nale\ifmmode~\dot{z}\else \.{z}\fi{}yty, P.~Horodecki, and J.~K. Korbicz.
\newblock ``System information propagation for composite structures''.
\newblock \href{https://dx.doi.org/10.1103/PhysRevA.98.022124}{Phys. Rev. A {\bf 98}, 022124}~(2018).

\bibitem{PhysRevLett.123.140402}
T.~K. Unden, D.~Louzon, M.~Zwolak, W.~H. Zurek, and F.~Jelezko.
\newblock ``Revealing the emergence of classicality using nitrogen-vacancy centers''.
\newblock \href{https://dx.doi.org/10.1103/PhysRevLett.123.140402}{Phys. Rev. Lett. {\bf 123}, 140402}~(2019).

\bibitem{Entropy2021_23(8)_995}
Barış Çakmak, Özgür~E. Müstecaplıoğlu, Mauro Paternostro, Bassano Vacchini, and Steve Campbell.
\newblock ``Quantum darwinism in a composite system: Objectivity versus classicality''.
\newblock \href{https://dx.doi.org/10.3390/e23080995}{Entropy{\bf 23}}~(2021).

\bibitem{PhysRevLett.128.010401}
Akram Touil, Bin Yan, Davide Girolami, Sebastian Deffner, and Wojciech~Hubert Zurek.
\newblock ``Eavesdropping on the decohering environment: Quantum darwinism, amplification, and the origin of objective classical reality''.
\newblock \href{https://dx.doi.org/10.1103/PhysRevLett.128.010401}{Phys. Rev. Lett. {\bf 128}, 010401}~(2022).

\bibitem{bartlett2007}
Stephen~D. Bartlett, Terry Rudolph, and Robert~W. Spekkens.
\newblock ``Reference frames, superselection rules, and quantum information''.
\newblock \href{https://dx.doi.org/10.1103/RevModPhys.79.555}{Rev. Mod. Phys. {\bf 79}, 555--609}~(2007).

\bibitem{quantumframes}
Flaminia Giacomini, Esteban Castro-Ruiz, and Časlav Brukner.
\newblock ``Quantum mechanics and the covariance of physical laws in quantum reference frames''.
\newblock \href{https://dx.doi.org/10.1038/s41467-018-08155-0}{Nature Communications{\bf 10}}~(2019).

\bibitem{bagan}
E.~Bagan, M.~Baig, and R.~Mu\~noz Tapia.
\newblock ``Aligning reference frames with quantum states''.
\newblock \href{https://dx.doi.org/10.1103/PhysRevLett.87.257903}{Phys. Rev. Lett. {\bf 87}, 257903}~(2001).

\bibitem{Chiribella_2004}
G.~Chiribella, Giacomo D'Ariano, P.~Perinotti, and M~Sacchi.
\newblock ``Efficient use of quantum resources for the transmission of a reference frame''.
\newblock \href{https://dx.doi.org/10.1103/PhysRevLett.93.180503}{Physical review letters {\bf 93}, 180503}~(2004).

\bibitem{Holevo1982}
Alexander Holevo.
\newblock ``Probabilistic and statistical aspects of quantum theory''.
\newblock \href{https://dx.doi.org/10.1007/978-88-7642-378-9}{Publications of the Scuola Normale Superiore}. Edizioni della Normale. Pisa~(2011).

\bibitem{Chiribella_2004_inny}
Giulio Chiribella, Giacomo~Mauro D’Ariano, Paolo Perinotti, and Massimiliano~F. Sacchi.
\newblock ``Covariant quantum measurements that maximize the likelihood''.
\newblock \href{https://dx.doi.org/10.1103/physreva.70.062105}{Physical Review A{\bf 70}}~(2004).

\bibitem{leftregular}
Anthony~W. Knapp.
\newblock ``Representation theory of semisimple groups: An overview based on examples ({PMS}-36)''.
\newblock \href{https://dx.doi.org/doi:10.1515/9781400883974}{Princeton University Press}. Princeton~(1986).

\bibitem{fuchs}
C.A. Fuchs and J.~van~de Graaf.
\newblock ``Cryptographic distinguishability measures for quantum-mechanical states''.
\newblock \href{https://dx.doi.org/10.1109/18.761271}{IEEE Transactions on Information Theory {\bf 45}, 1216--1227}~(1999).

\bibitem{Nielsen_Chuang_2010}
Michael~A. Nielsen and Isaac~L. Chuang.
\newblock ``Quantum computation and quantum information: 10th anniversary edition''.
\newblock \href{https://dx.doi.org/10.1017/CBO9780511976667}{Cambridge University Press}. ~(2010).

\bibitem{Klein}
Felix Klein.
\newblock ``Vergleichende betrachtungen {\"u}ber neuere geometrische forschungen''.
\newblock \href{https://dx.doi.org/10.1007/BF01446615}{Mathematische Annalen {\bf 43}, 63--100}~(1893).

\bibitem{Korbicz2017}
J.~K. Korbicz, E.~A. Aguilar, P.~\ifmmode \acute{C}\else \'{C}\fi{}wikli\ifmmode~\acute{n}\else \'{n}\fi{}ski, and P.~Horodecki.
\newblock ``Generic appearance of objective results in quantum measurements''.
\newblock \href{https://dx.doi.org/10.1103/PhysRevA.96.032124}{Phys. Rev. A {\bf 96}, 032124}~(2017).

\bibitem{tuziemski2015}
J.~Tuziemski and J.~K. Korbicz.
\newblock ``Dynamical objectivity in quantum brownian motion''.
\newblock \href{https://dx.doi.org/10.1209/0295-5075/112/40008}{Europhysics Letters {\bf 112}, 40008}~(2015).

\bibitem{PhysRevResearch.3.033148}
Carlo~Maria Scandolo, Roberto Salazar, Jaros\l{}aw~K. Korbicz, and Pawe\l{} Horodecki.
\newblock ``Universal structure of objective states in all fundamental causal theories''.
\newblock \href{https://dx.doi.org/10.1103/PhysRevResearch.3.033148}{Phys. Rev. Res. {\bf 3}, 033148}~(2021).

\bibitem{bib_tridiagonal_toeplitz_matrices}
Silvia Noschese, Lionello Pasquini, and Lothar Reichel.
\newblock ``Tridiagonal toeplitz matrices: Properties and novel applications''.
\newblock \href{https://dx.doi.org/10.1002/nla.1811}{Numerical Linear Algebra with Applications{\bf 20}}~(2013).

\bibitem{bib_charakter_su(2)}
Brian Hall.
\newblock ``The compact group approach to representation theory''.
\newblock \href{https://dx.doi.org/10.1007/978-3-319-13467-3_12}{Pages 343--370}.
\newblock Springer International Publishing. Cham~(2015).

\bibitem{lsuumamtionofseries54trhh3}
``Trigonometry/{T}he summation of finite series''.
\newblock \url{https://en.wikibooks.org/wiki/Trigonometry/The_summation_of_finite_series}.
\newblock Accessed: 2025-05-31.

\end{thebibliography}

\onecolumn
\appendix

\section{The \texorpdfstring{\( \mathrm{SU(2)} \)}{SU(2)} Group and Its Representations}
\label{appendix:su2group}
The \(\mathrm{SU(2)}\) group is the group of \(2 \times 2\) unitary matrices with determinant \(1\), and the group operation of matrix multiplication. Let \(J_x\), \(J_y\), and \(J_z\) be generators of the Lie algebra \(\mathfrak{su}(2)\). In the fundamental representation we have
\begin{align}
    J_x &= \frac{1}{2} \sigma_x = \frac{1}{2}
    \begin{bmatrix}
    0 & 1 \\
    1 & 0
    \end{bmatrix} \, ,\\
    J_y &= \frac{1}{2} \sigma_y = \frac{1}{2}
    \begin{bmatrix}
    0 & -\mathrm{i} \\
    \mathrm{i} & 0
    \end{bmatrix} \, ,\\
    J_z &= \frac{1}{2} \sigma_z = \frac{1}{2}
    \begin{bmatrix}
    1 & 0 \\
    0 & -1
    \end{bmatrix} \, .
\end{align}
The Casimir operator equals to
\begin{align}
    J^2 = J_x^2+J_y^2+J_z^2 = \frac{3}{4}
    \begin{bmatrix}
    1 & 0 \\
    0 & 1
    \end{bmatrix} \, .
\end{align}
The \(\mathrm{SU(2)}\) group is diffeomorphic to the 3-sphere \( S^3 \). Therefore, it can be parametrized by the three hyperspherical angles \(\theta\), \(\psi\) and \(\varphi\) in the following way:
\begin{align}\label{ty645768Yhyjgkjk677}
    g(\theta, \psi, \varphi) =
    \begin{bmatrix}
    x_1 + \mathrm{i} x_2 & x_3 + \mathrm{i} x_4\\
    - x_3 + \mathrm{i} x_4 & x_1 - \mathrm{i} x_2
    \end{bmatrix} \, ,
\end{align}
where
\begin{align}
    x_1 = & \cos \theta \, , \label{rtygujhHGJ789gsF1}\\
    x_2 = & \sin \theta \cos \psi \, , \label{rtygujhHGJ789gsF2}\\
    x_3 = & \sin \theta \sin \psi \cos \varphi \, , \label{rtygujhHGJ789gsF3}\\
    x_4 = & \sin \theta \sin \psi \sin \varphi \, \label{rtygujhHGJ789gsF4}.
\end{align}
The hyperspherical coordinates take values within the intervals \(0 \leq \theta \leq \pi\), \(0 \leq \psi \leq \pi\), and \(0\leq \varphi \leq 2 \pi\). We denote the normalized Haar measure on the \( \mathrm{SU(2)} \) group by \( \mu_{\mathrm{SU(2)}} \). The following formula holds:
\begin{align}\label{njcxdjncdxjnhcdxnhjiuwf}
    \begin{split}
        \int\displaylimits_{\mathrm{SU(2)}} f(g) \, \mu_{\mathrm{SU(2)}}(\mathrm{d}g) = \frac{1}{2 \pi^2} \int_{0}^{\pi} \sin^2 \theta \, \mathrm{d}\theta \int_{0}^{\pi} \sin \psi \, \mathrm{d}\psi \int_{0}^{2 \pi} \mathrm{d}\varphi \, f \left[ g(\theta,\psi,\varphi) \right] \, .
    \end{split}
\end{align}
When we integrate over the \( \mathrm{SU(2)} \) group, we will always do so with respect to the Haar measure \( \mu_{\mathrm{SU(2)}} \); therefore, we will usually omit its symbol. As a side note, let us notice that for every \( g \in \mathrm{SU(2)} \) we have
\begin{align}\label{rdbtfygj4576nhTJ32c0}
    \Tr ( g ) = \Tr
    \begin{bmatrix}
    x_1 + \mathrm{i} x_2 & x_3 + \mathrm{i} x_4\\
    - x_3 + \mathrm{i} x_4 & x_1 - \mathrm{i} x_2
    \end{bmatrix}
    = 2 x_1 = \Tr
    \begin{bmatrix}
    x_1 - \mathrm{i} x_2 & - x_3 - \mathrm{i} x_4\\
    x_3 - \mathrm{i} x_4 & x_1 + \mathrm{i} x_2
    \end{bmatrix}
    = \Tr (g^{-1})\, .
\end{align}

The \(\mathrm{SO(3)}\) group is the group of \(3 \times 3\) orthogonal matrices with determinant \(1\), and the group operation of matrix multiplication. By the usual construction, we can introduce the two-to-one and surjective homomorphism
\begin{align}
    \mathcal{R}: \mathrm{SU(2)} \to \mathrm{SO(3)} \, ,
\end{align}
which for every \( g \in \mathrm{SU(2)} \) has the following property
\begin{align}\label{hytjyh5587678hjjYUJHG}
    \mathcal{R}(g) = \mathcal{R}(-g) \, .
\end{align}
It can be easily demonstrated (see Eqs. \eqref{ty645768Yhyjgkjk677} -- \eqref{rtygujhHGJ789gsF4}) that
\begin{align}\label{rertgfhy467878GFuy}
    g(\theta,\psi,\varphi) = -g(\pi - \theta, \pi - \psi, \varphi + \pi) \, ,
\end{align}
thus
\begin{align}\label{56gtyuhHG67gh6oNJH}
    \mathcal{R} \left[ g(\theta,\psi,\varphi) \right] = \mathcal{R} \left[ g(\pi - \theta, \pi - \psi, \varphi + \pi) \right] \, .
\end{align}
We denote the normalized Haar measure on the \( \mathrm{SO(3)} \) group by \( \mu_{\mathrm{SO(3)}} \). It is a known fact that for every function \( f \in \mathcal{L}^1 (\mathrm{SO(3)},\mu_{\mathrm{SO(3)}}) \) we have
\begin{align}\label{yvgjbh547687bcfgVGHJNH7d}
    \int\displaylimits_{\mathrm{SU(2)}} f \left[ \mathcal{R}(g) \right] \, \mu_{\mathrm{SU(2)}}(\mathrm{d}g) = \int\displaylimits_{\mathrm{SO(3)}} f(R) \, \mu_{\mathrm{SO(3)}}(\mathrm{d}R) \, .
\end{align}
When we integrate over the \( \mathrm{SO(3)} \) group, we will always do so with respect to the Haar measure \( \mu_{\mathrm{SO(3)}} \); therefore, we will also omit its symbol.

We denote by \( U_j \) the irreducible representation of the \( \mathrm{SU(2)} \) group with spin \( j \). It acts on the Hilbert space \( \mathcal{H}_{j} \), in which we introduce the orthonormal basis consisting of the following vectors
\begin{align}
    \left\{ |j, m\rangle \big| m = -j, \ldots, j \right\} \, ,
\end{align}
where \( |j, m\rangle \) is an eigenvector with eigenvalue \(m\) of the operator corresponding to the element \(J_z\) of the Lie algebra \( \mathfrak{su}(2) \). We introduce the following notation for the diagonal matrix elements of the representations:
\begin{align}\label{appfre9tgtg)HnjJmm}
    D^j_{m, m} (g) = \langle j, m | U_j (g) | j, m \rangle \, .
\end{align}

\section{Overview of Chiribella et al.'s Protocol for the Transmission of a Reference Frame}
\label{appendix:chiribella}
This appendix presents a comprehensive summary of the key results and concepts from \cite{Chiribella_2004}, focusing specifically on the aspects relevant to our study. By providing a self-contained overview, we aim to clarify the theoretical tools and assumptions developed by the authors, which constitute part of the foundation upon which our analysis is built.

From a mathematical point of view, the state of \( N \) spins \( \frac{1}{2}\) is described by a normalized vector in the Hilbert space
\begin{align}\label{appmk9nbfrdcdry48}
    \mathcal{H}_N = \bigotimes_{n=1}^{N} \mathcal{H}_{\frac{1}{2}} \, .
\end{align}
On the Hilbert space \( \mathcal{H}_N \) acts the following representation of the \( \mathrm{SU(2)} \) group:
\begin{align}
    U = \bigotimes_{n=1}^{N} U_\frac{1}{2} \, .
\end{align}
\( \mathcal{H}_N \) can be decomposed into an orthogonal direct sum of subspaces
which are irreducible under the action of the \( \mathrm{SU(2)} \) group, i.e.
\begin{align}
    \mathcal{H}_N = \bigotimes_{n=1}^{N} \mathcal{H}_{\frac{1}{2}} = \bigoplus_{j=j_0}^{J} \bigoplus_{\alpha=1}^{n_j} \mathcal{H}_{j \alpha} \, ,
\end{align}
where \( j \) is the quantum number of the total angular momentum, \( j_0 = 0\) for \(N\) even and \( j_0 = \frac{1}{2}\) for \(N\) odd, \( J = \frac{N}{2}\), \( \alpha \) labels different equivalent representations with the same \( j \), and \( n_j \) is the number of equivalent representations with spin \( j \). It is given by
\begin{align}\label{nj}
    n_j = \frac{2j + 1}{J + j + 1} \binom{2J}{J + j} \, .
\end{align}
In each invariant subspace \( \mathcal{H}_{j \alpha} \) we introduce the orthonormal basis consisting of the following vectors:
\begin{align}
    \left\{ |j\alpha, m \rangle | \, m = -j, \ldots, j \right\} \, ,
\end{align}
where \( |j\alpha, m\rangle \) is an eigenvector with eigenvalue \(m\) of the operator corresponding to the element \(J_z\) of the Lie algebra \( \mathfrak{su}(2) \). We will say that two vectors \( |\psi_{j \alpha} \rangle \in \mathcal{H}_{j \alpha} \) and \( |\varphi_{j \beta} \rangle \in \mathcal{H}_{j \beta} \) are iso-orthogonal if
\begin{align}
\langle \psi_{j \alpha} | \left( \sum_{m=-j}^{j} |j \alpha, m\rangle \langle j \beta, m| \right) | \varphi_{j \beta} \rangle = 0 \, .
\end{align}
We introduce Alice's reference state \( | A \rangle \in \mathcal{H}\) and denote 
\begin{align}\label{app02jdn3n3idn3di9m2sz}
    | A(g_0) \rangle = U(g_0) | A \rangle \, .
\end{align}
The state \( | A \rangle \) will be constructed in order to communicate the reference frame as efficiently as possible. To do this, we have to use as many equivalent representations as possible. The key insight is that the maximum number of useful representations within class \( j \) is not \( n_j \), but \( k_j = \min\{n_j, 2j + 1\} \). This reflects the fact that equivalent representations are beneficial only when iso-orthogonal vectors in different subspaces \( \mathcal{H}_{j \alpha} \) are used (see \cite{Chiribella_2004_inny}). It can easily be seen that
\begin{align}
    k_j =
    \begin{cases} 
    2j + 1  & j_0 \leq j < J\\
    1     & j = J 
    \end{cases} \, .
\end{align}
We choose the following Alice’s reference vector:
\begin{align}\label{appstsrgu6ygr67i3446j}
    | A \rangle = \sum_{j=j_0}^{J-1} \sum_{\alpha=1}^{2j+1} \frac{A_j}{\sqrt{2j+1}} | j \alpha, m(\alpha) \rangle \, ,
\end{align}
where \( m(\alpha) \) is an injective function of \( \alpha \) (we have to take iso-orthogonal vectors in different equivalent subrepresentations). Without loss of generality we can take \( A_j \geq 0 \). We have arbitrarily set the component of \( | A \rangle \) in \( \mathcal{H}_{J 1} \) to zero, because in the asymptotic limit of large \(N\) the contribution of this representation becomes negligible (see \cite{Chiribella_2004}). Given that our analysis concerns the asymptotic limit \( N \to \infty \), this simplification is justified. State \( | A \rangle \) has to be normalized, so
\begin{align}
    \langle A | A \rangle = \sum_{j=j_0}^{J-1} \left| A_j \right|^2 = 1 \, .
\end{align}
Bob wants to extract rotation \(\mathcal{R}(g_0)\) from the state \( | A (g_0) \rangle \). To do this, he can perform measurements on this state. The most general measurement is described by a POVM, i.e., by a set of positive operators \( \{ M(g') \}_{g' \in \mathrm{SU(2)}} \) on the Hilbert space \( \mathcal{H}_N \) satisfying condition
\begin{align}
    \int_{\mathrm{SU(2)}} \mathrm{d}g' M(g') = \mathbb{1} \, ,
\end{align}
where \( \mathrm{d}g' \) is the normalized Haar measure on the \( \mathrm{SU(2)} \) group. The probability density of estimating \( g' \) when the measured state is \( | A (g_0) \rangle \) is given by the Born rule and equals
\begin{align}
    p(g'|g_0) = \mathrm{Tr} \left[ M(g')|A(g_0) \rangle \langle A(g_0)| \right] \, .
\end{align}
The efficiency of a strategy is characterized by the transmission error
\begin{align}
    e(g', g_0) = \sum_{\alpha \in  \{ x, y, z \}} \left| \mathcal{R}(g') \hat{e}_\alpha - \mathcal{R}(g_0) \hat{e}_\alpha \right|^2,
\end{align}
Maximizing efficiency is equivalent to minimizing the average error
\begin{align}\label{appw983hd32m923dk20dA}
    \langle e \rangle = \int_{\mathrm{SU(2)}} \mathrm{d}g_0 \int_{\mathrm{SU(2)}} \mathrm{d}g' p(g' | g_0) e(g', g_0) \, .
\end{align}
In the above formula we assume a uniform a priori distribution for \(g_0\), reflecting the fact that \( g_0 \) is entirely unknown. Let us note that
\begin{align}
    e(g', g_0) = e (hg', hg_0)
\end{align}
for every \( h \in \mathrm{SU(2)} \). Therefore, we can assume that Bob's measurement strategy is described by a covariant POVM, i.e.
\begin{align}
    M(g') = U(g') \Xi U^{\dagger}(g') \, ,
\end{align}
where \( \Xi \) is a positive operator on \( \mathcal{H}_N \). Next, we have to determine the specific covariant POVM that Bob should use to extract \(g_0\) from the state \( |A(g_0) \rangle \). It is easy to see that the Alice’s reference vector \( |A \rangle \) (Eq. \eqref{appstsrgu6ygr67i3446j}) lies in the following invariant subspace of \( \mathcal{H}_N \):
\begin{align}
    \mathcal{K} = \bigoplus_{j=j_0}^{J-1} \bigoplus_{\alpha = 1}^{2j+1} \mathcal{H}_{j \alpha} \, .
\end{align}
Therefore, the probability distribution
\begin{align}\label{appAspSE932h39dn}
\begin{split}
    p(g' | g_0) &= \mathrm{Tr} \left[ M(g')|A(g_0) \rangle \langle A(g_0)| \right] =\\
    &= \langle A(g_0) | U(g') \Xi U^{\dagger}(g') | A(g_0) \rangle
\end{split}
\end{align}
depends only on the restriction of \( \Xi \) to the invariant subspace \( \mathcal{K} \), i.e. it depends only on the operator
\begin{align}
    \xi = P_\mathcal{K} \Xi P_\mathcal{K} \, ,
\end{align}
where \( P_\mathcal{K} \) is the orthogonal projection on the subspace \( \mathcal{K} \). Instead of optimizing Bob’s POVM to minimize the average error (Eq. \eqref{appw983hd32m923dk20dA}), we use the maximum likelihood POVM (the one that maximizes the peak \( p(g_0 | g_0) \) in the probability distribution \( p(g' | g_0) \)). In this case we have (see \cite{Chiribella_2004_inny})
\begin{align}
    \xi = | B \rangle \langle B| \, ,
\end{align}
where
\begin{align}\label{appmdj7suhe56wyegds2}
    | B \rangle = \sum_{j=j_0}^{J-1} \sum_{\alpha=1}^{2j+1} \sqrt{2j+1} \big| \, j \alpha, m(\alpha) \rangle \, .
\end{align}
It is easy to see that now Eq. \eqref{appAspSE932h39dn} takes the form
\begin{align}
    p(g' | g_0) = \left| \langle A(g_0) | B(g') \rangle \right|^2 \, ,
\end{align}
where
\begin{align}\label{appmx39xnj39v548z20fafU}
    | B(g') \rangle = U(g') | B \rangle \, .
\end{align}
We also have to find the optimal coefficients \( \{ A_j \}\) in the Alice's reference vector (Eq. \eqref{appstsrgu6ygr67i3446j}). It was shown in \cite{bagan} that minimizing the average error \( \langle e \rangle \) is equivalent to maximizing the average character
\begin{align}
    \langle \chi \rangle = \int_{\mathrm{SU(2)}} \mathrm{d}g \, \chi(g)  p(g | \mathbb{1}_2) \, ,
\end{align}
where \(\mathbb{1}_2 \in \mathrm{SU(2)} \) is the identity element,
\begin{align}
    \chi(g) = \sum_{m=-1}^{1} D^1_{m,m}(g) \, ,
\end{align}
and \( D^j_{m, m} \left( g \right) \) are the diagonal matrix elements defined by Eq. \eqref{appfre9tgtg)HnjJmm}. It was shown in \cite{Chiribella_2004} that the average character is given by the following formula
\begin{align}
    \langle \chi \rangle = \sum_{j=j_0}^{J-1} \sum_{j'=j_0}^{J-1} A_j M_{j j'} A_{j'} = \mathbf A^T M \mathbf A \, ,
\end{align}
where \( M \) is the following tridiagonal matrix:
\begin{align}
    M &=
    \begin{bmatrix}
    \zeta & 1 & 0 & 0 & \dots  & 0 \\
    1 & 1 & 1 & 0 & \dots  & 0 \\
    0 & 1 & 1 & 1 & \dots  & 0 \\
    0 & 0 & 1 & 1 & \dots  & 0 \\
    \vdots & \vdots & \vdots & \vdots & \ddots & \vdots \\
    0 & 0 & 0 & 0 & \dots  & 1
    \end{bmatrix} \, ,\\
    \mathbf A^T &=
    \begin{bmatrix}
    A_{j_0} & A_{j_0+1} & \ldots & A_{J-1} \\
    \end{bmatrix} \, ,
\end{align}
and \( \zeta = 0(1) \) for even (odd) values of \( N \). We have omitted the row and column corresponding to  \( \mathcal{H}_{J 1} \) due to our assumption that \( A_J=0 \). Let us note that the size of the matrix \( M \) is \( n \times n\), where
\begin{align}\label{appericn39cn39cn33}
    n = J - j_0 =
    \begin{cases} 
    J - \frac{1}{2}    & \mathrm{if} \; 2 \nmid N \\
    J                  & \mathrm{if} \; 2 \mid N 
    \end{cases} \, .
\end{align}
Since the vector \(\mathbf A\) is normalized, the maximum \( \langle \chi \rangle \) is simply equal to the largest eigenvalue of the matrix \( M \), and the optimal coefficients \( \{ A_j \} \) are the components of the corresponding normalized eigenvector \(\mathbf A\).

\section{Tridiagonal Toeplitz Matrices}
\label{appendix:ttm}
This appendix is dedicated to finding a normalized eigenvector corresponding to the largest eigenvalue of the following \(n \times n\) tridiagonal Toeplitz matrix:
\begin{align}\label{ttm}
    T_n :=
    \begin{bmatrix}
    1 & 1 & 0 & \dots  & 0 \\
    1 & 1 & 1 & \dots  & 0 \\
    0 & 1 & 1 & \dots  & 0 \\
    \vdots & \vdots & \vdots & \ddots & \vdots \\
    0 & 0 & 0 & \dots  & 1
    \end{bmatrix} \, .
\end{align}
It has the following eigenvalues (see \cite{bib_tridiagonal_toeplitz_matrices}):
\begin{align}
    \lambda_r = 1 + 2 \cos \left (\frac{r\pi}{n+1} \right ) \, ,
\end{align}
where \(r \in \overline{1,n}\). The greatest eigenvalue is
\begin{align} \label{tge}
    \lambda := \lambda_1 = 1 + 2 \cos \left (\frac{\pi}{n+1} \right ) \, .
\end{align}
We want to find a normalized eigenvector of the matrix \(T_n\) corresponding to the eigenvalue \( \lambda \). We denote it in the following way:
\begin{align}
    v = \begin{bmatrix}
    v^1 \\
    v^2 \\
    \vdots \\
    v^n 
\end{bmatrix} \, .
\end{align}
We have (see \cite{bib_tridiagonal_toeplitz_matrices})
\begin{align}
    v^l = C_n \sin \left( \frac{l\pi}{n+1} \right) \, ,
\end{align}
where \(l \in \overline{1,n}\) and \(C_n\) is a normalizing constant. We have
\begin{align}
\begin{split}
    1 &= || v ||^2 = \sum_{l=1}^{n} | v^l |^2 = \sum_{l=1}^{n} \left| C_n \right|^2 \sin^2 \left( \frac{l\pi}{n+1} \right) = \left| C_n \right|^2 \sum_{l=1}^{n} \frac{1}{2} \left( 1 - \cos \left( \frac{2l\pi}{n+1}\right) \right) =\\
    &= \frac{\left| C_n \right|^2}{2} \left( \sum_{l=1}^{n} 1 - \sum_{l=1}^{n} \cos \left( \frac{2l\pi}{n+1} \right) \right) = \frac{\left| C_n \right|^2}{2} \left( n - \sum_{l=1}^{n+1} \cos \left( \frac{2l\pi}{n+1} \right) + \cos 2\pi \right) = \left| C_n \right|^2 \frac{n+1}{2} \, ,\\
\end{split}
\end{align}
where we used the fact that
\begin{align}
    \sum_{l=1}^{n+1} \cos \left( \frac{2l\pi}{n+1} \right) = \Re \sum_{l=1}^{n+1} \exp \left( \frac{2l\pi\mathrm{i}}{n+1} \right) = \Re \sum_{l=1}^{n+1} \left[ \exp \left( \frac{2\pi\mathrm{i}}{n+1} \right) \right]^l = 0 \, .
\end{align}
This means that
\begin{align}\label{jbwiwejksjdgftippmsafxebwy}
    \left| C_n \right| = \sqrt{\frac{2}{n+1}} \, ,
\end{align}
and we can take
\begin{align}\label{w0dn3dn2m0cm94b3bcn3}
    v^l = \sqrt{\frac{2}{n+1}} \sin \left( \frac{l\pi}{n+1} \right) \, ,
\end{align}
where \(l \in \overline{1,n}\).

\section{Calculation of the Scalar Product \texorpdfstring{\( \langle A \left( g \right) | B \left( g' \right) \rangle \)}{⟨A(g)|B(g′)⟩}}
\label{appendix:scalarproduct}
Our main goal in this appendix is to calculate the scalar product \( \langle A \left( g \right) | B \left( g' \right) \rangle \). Alice's reference vector is as follows (see Eq. \eqref{stsrgu6ygr67i3446j}):
\begin{align}\label{m4x3c455678cGRTFdRTRDf}
    | A \rangle = \sum_{j=j_0}^{J-1} \sum_{\alpha=1}^{2j+1} \frac{A_j}{\sqrt{2j+1}} | j \alpha, m(\alpha) \rangle \, .
\end{align}
The coefficients \(A_j\), where \(j \in \overline{j_0, J-1}\), are given by the following formula (see Eqs. \eqref{ericn39cn39cn33}, \eqref{njsuehdnbhhg2938uwj}, \eqref{9phOVCFGHa2gv} and \eqref{nwio4h93nc3a}):
\begin{align}\label{mj2h37ing374fnc4fhrfvfa}
    A_j = v^{j-j_0+1} = \sqrt{\frac{2}{n+1}} \sin \left( \frac{j-j_0+1}{n+1} \pi \right) \, ,
\end{align}
where
\begin{align}\label{bhj7bUYh87h01bxuiy78hAd}
n = J - j_0 \, .
\end{align}
For the above formulas to make sense, we have to assume that \( J \geq 1 \). The state \( | A \rangle \) is normalized. Indeed, we have
\begin{align}\label{yfyghv7btuyh6uv7rbyghFFJn7}
\langle A | A \rangle = \sum_{j=j_0}^{J-1} \sum_{\alpha=1}^{2j+1} \frac{\left| A_j \right|^2}{2j+1} = \sum_{j=j_0}^{J-1} \left| A_j \right|^2 = \sum_{l=1}^{n} | v^l |^2 = 1 \, ,
\end{align}
where the last equality was proven in Appendix \ref{appendix:ttm}. Equation \eqref{mdj7suhe56wyegds2} tells us that
\begin{align}\label{mnf8bv57c44TvgGTYbYT12o}
    | B \rangle = \sum_{j=j_0}^{J-1} \sum_{\alpha=1}^{2j+1} \sqrt{2j+1} \big| \, j \alpha, m(\alpha) \rangle \, .
\end{align}
The scalar product \( \langle A \left( g \right) | B \left( g' \right) \rangle \) is given by the following expression (see Eqs. \eqref{02jdn3n3idn3di9m2sz}, \eqref{mx39xnj39v548z20fafU}, \eqref{m4x3c455678cGRTFdRTRDf} and \eqref{mnf8bv57c44TvgGTYbYT12o}):
\begin{align}\label{ghbgiuyt6756tftfg}
\begin{split}
    \langle A(g) | B(g') \rangle &= \langle A | U(g^{-1}g') | B \rangle = \sum_{j=j_0}^{J-1} A_j \sum_{\alpha=1}^{2j+1} \langle j \alpha, m(\alpha) | U(g^{-1}g') | j \alpha, m(\alpha) \rangle =\\
    &= \sum_{j=j_0}^{J-1} A_j \sum_{\alpha=1}^{2j+1} D^j_{m(\alpha), m(\alpha)}(g^{-1}g') = \sum_{j=j_0}^{J-1} A_j \sum_{m=-j}^{j} D^j_{m, m}(g^{-1}g') = (\star) \, ,
\end{split}
\end{align}
where \( D^j_{m, m} \) are the matrix elements of the irreducible representations of the \(\mathrm{SU(2)}\) group (described in detail in Appendix \ref{appendix:su2group}, see in particular Eq. \eqref{appfre9tgtg)HnjJmm}) and \(m(\alpha)\) is an injective function of \(\alpha\). The last sum is just the character of the \(\mathrm{SU(2)}\) group representation which equals to (see \cite{bib_charakter_su(2)})
\begin{align}\label{cha98hg7BGHJJjhnU8has}
    \sum_{m=-j}^{j} D^j_{m, m}(g^{-1}g') = \frac{\sin \left[ (2j+1)\theta \right]}{\sin \theta} \, ,
\end{align}
where \(\theta \in [0,\pi]\) corresponds to the group element \(g^{-1}g' \in \mathrm{SU(2)}\). It is given by the following equation:
\begin{align}
    \cos\theta = \frac{1}{2} \Tr \left( g^{-1}g' \right) \, .
\end{align}
It is easy to see that  \( \theta = 0 \iff g = g' \) and \( \theta = \pi \iff g = -g' \). Appendix \ref{appendix:su2group} provides a detailed explanation of the \(\mathrm{SU(2)}\) group parametrization using hyperspherical coordinates. We substitute Formula \eqref{cha98hg7BGHJJjhnU8has} into Eq. \eqref{ghbgiuyt6756tftfg} and obtain
\begin{align}
    (\star) = \sum_{j=j_0}^{J-1} A_j \frac{\sin \left[ (2j+1)\theta \right]}{\sin \theta} = \sqrt{\frac{2}{n+1}} \frac{1}{\sin \theta} \sum_{j=j_0}^{J-1} \sin \left( \frac{j-j_0+1}{n+1} \pi \right) \sin \left[ (2j+1)\theta \right] = (\star \star) \, ,
\end{align}
where we have also substituted \(A_j\) given by Formula \eqref{mj2h37ing374fnc4fhrfvfa}. Subsequently, we use the trigonometric identity \( 2\sin \alpha \sin \beta = \cos(\alpha-\beta) - \cos(\alpha+\beta)\) and obtain
\begin{align}\label{nc93fhnu933ni3xjn3jG7}
\begin{split}
    (\star \star) &= \sqrt{\frac{2}{n+1}} \frac{1}{\sin \theta}\sum_{j=j_0}^{J-1} \frac{1}{2} \left[ \cos \left( \frac{j-j_0+1}{n+1} \pi - (2j+1)\theta \right)  - \cos \left( \frac{j-j_0+1}{n+1} \pi + (2j+1)\theta \right) \right] =\\
    &= \sqrt{\frac{2}{n+1}} \frac{1}{\sin \theta }\sum_{j=j_0}^{J-1} \frac{1}{2} \left[ \cos \left( c_- + j d_- \right)  - \cos \left( c_+ + j d_+ \right) \right] = (\circ) \, ,
\end{split}
\end{align}
where
\begin{align}
    c_{\pm} = \frac{1-j_0}{n+1} \pi \pm \theta
\end{align}
and
\begin{align}\label{niwdmbermhwu823h}
    d_{\pm} = \frac{\pi}{n+1} \pm 2\theta \, .
\end{align}
For \(n \in \mathbb{Z}^+\) we have the following summation formula (see \cite{lsuumamtionofseries54trhh3}):
\begin{align}
    \sum_{r=0}^{n-1} \cos \left( a + rb \right) = \frac{\sin \left( \frac{1}{2} nb \right)}{\sin \left( \frac{1}{2}b \right)} \cos \left[ a + (n-1) \frac{b}{2} \right] \, .
\end{align}
In our case this gives us
\begin{align}
    \sum_{j=j_0}^{J-1} \cos \left( c + jd \right) = \sum_{r=0}^{J-j_0-1} \cos \left( c + j_0 d + rd \right) = \frac{\sin \left[ \frac{1}{2} (J-j_0)d \right]}{\sin \left( \frac{1}{2}d \right)} \cos \left[ c + j_0 d + (J-j_0-1) \dfrac{d}{2} \right] \, .
\end{align}
We substitute the above formula into Eq. \eqref{nc93fhnu933ni3xjn3jG7} and obtain
\begin{align}
\begin{split}
    (\circ) = \frac{1}{2} \sqrt{\frac{2}{n+1}} \frac{1}{\sin \theta } & \left[ \frac{\sin \left( \frac{1}{2} (J-j_0)d_- \right)}{\sin \left( \frac{1}{2}d_- \right)} \cos \left( c_- + (J+j_0-1) \dfrac{d_-}{2} \right) \right. +\\
    &- \left. \frac{\sin \left( \frac{1}{2} (J-j_0)d_+ \right)}{\sin \left( \frac{1}{2}d_+ \right)} \cos \left( c_+ + (J+j_0-1) \dfrac{d_+}{2} \right) \right] =  (\circ \circ) \, .
\end{split}
\end{align}
Let us note that
\begin{align}
    c_\pm + (J+j_0-1) \dfrac{d_\pm}{2} = \dfrac{\pi}{2} \pm (J+j_0) \theta \, ,
\end{align}
therefore
\begin{align}
\begin{split}
    (\circ \circ) &= \frac{1}{2} \sqrt{\frac{2}{n+1}} \frac{\sin \left[ (J+j_0) \theta \right]}{\sin \theta } \left[ \dfrac{\sin \left( \frac{1}{2} (J-j_0)d_- \right)}{\sin \left( \frac{1}{2}d_- \right)} + \dfrac{\sin \left( \frac{1}{2} (J-j_0)d_+ \right)}{\sin \left( \frac{1}{2}d_+ \right)} \right] =\\
    &= \frac{1}{2} \sqrt{\frac{2}{n+1}} \frac{\sin \left[ (J+j_0) \theta \right]}{\sin \theta } \dfrac{\sin \left( \frac{1}{2} n d_- \right) \sin \left( \frac{1}{2}d_+ \right) + \sin \left( \frac{1}{2} n d_+ \right) \sin \left( \frac{1}{2}d_- \right)}{\sin \left( \frac{1}{2}d_+ \right) \sin \left( \frac{1}{2}d_- \right)} =  (\diamond) \, .
\end{split}
\end{align}
Next, we again use the trigonometric identity \( 2\sin \alpha \sin \beta = \cos(\alpha-\beta) - \cos(\alpha+\beta) \), substitute \(d_\pm\) (Eq. \eqref{niwdmbermhwu823h}), simplify and obtain
\begin{align}
    (\diamond) = \frac{1}{2} \sqrt{\frac{2}{n+1}} \frac{\sin \left[ (J+j_0) \theta \right]}{\sin \theta } \dfrac{\cos \left[ \dfrac{n-1}{n+1} \dfrac{\pi}{2} - (n+1) \theta \right] + \cos \left[ \dfrac{n-1}{n+1} \dfrac{\pi}{2} + (n+1) \theta \right]}{\cos 2 \theta - \cos \left( \dfrac{\pi}{n+1} \right)} = (\diamond \diamond) \, .
\end{align}
Then we use the trigonometric identity \( \cos \alpha + \cos \beta =  2\cos\left[ \frac{1}{2}(\alpha + \beta) \right] \cos\left[ \frac{1}{2}(\alpha - \beta) \right] \), simplify and obtain
\begin{align}
\begin{split}
    (\diamond \diamond) &= \sqrt{\frac{2}{n+1}} \frac{\sin \left[ (J+j_0) \theta \right]}{\sin \theta} \dfrac{\cos \left( \dfrac{n-1}{n+1} \dfrac{\pi}{2} \right) \cos \left[ (n+1) \theta \right]}{\cos 2 \theta - \cos \left( \dfrac{\pi}{n+1} \right)} =\\
    &= \sqrt{\frac{2}{n+1}} \frac{\sin \left[ (J+j_0) \theta \right]}{\sin \theta} \dfrac{\sin \left( \dfrac{\pi}{n+1} \right) \cos \left[ (n+1) \theta \right]}{\cos 2 \theta - \cos \left( \dfrac{\pi}{n+1} \right)} \, ,
\end{split}
\end{align}
where we also used the identity \( \cos\left( \frac{\pi}{2} - \alpha \right) = \sin \alpha \). Finally, we substitute \(n=J-j_0\) (Eq. \eqref{bhj7bUYh87h01bxuiy78hAd}) into the above formula, which leads to
\begin{align}\label{7uhgtghhgytyu67ghgmWd}
    \langle A(g) | B(g') \rangle = \sqrt{\frac{2}{J-j_0+1}} \frac{\sin \left[ (J+j_0) \theta \right]}{\sin \theta} \dfrac{\sin \left( \dfrac{\pi}{J-j_0+1} \right) \cos \left[ (J-j_0+1) \theta \right]}{\cos 2 \theta - \cos \left( \dfrac{\pi}{J-j_0+1} \right)} \, ,
\end{align}
and at the end gives us the following conditional probability density function:
\begin{align}
    p(g' | g) = \left| \langle A(g) | B(g') \rangle \right|^2 = \frac{2}{J-j_0+1} \frac{\sin^2 \left[ (J+j_0) \theta \right]}{\sin^2 \theta } \dfrac{\sin^2 \left( \dfrac{\pi}{J-j_0+1} \right) \cos^2 \left[ (J-j_0+1) \theta \right]}{\left[ \cos  2 \theta - \cos \left( \dfrac{\pi}{J-j_0+1} \right) \right]^2} \, .
\end{align}
This is the most important formula derived in this appendix. It is worth noting that this is an exact result without approximation, given the definitions of \(|A\rangle\) and \(|B\rangle\) (Eqs. \eqref{m4x3c455678cGRTFdRTRDf}, \eqref{mj2h37ing374fnc4fhrfvfa} and \eqref{mnf8bv57c44TvgGTYbYT12o}). In the special case, when \(g=g'\), we have \(\theta=0\), and Eq. \eqref{7uhgtghhgytyu67ghgmWd} gives us
\begin{align}
        \langle A | B \rangle = \sqrt{\frac{2}{J-j_0+1}} (J+j_0) \dfrac{\sin \left( \dfrac{\pi}{J-j_0+1} \right)}{1 - \cos \left( \dfrac{\pi}{J-j_0+1} \right)} \, .
\end{align}
Although we will not use it in our main work, for completeness we have also calculated the scalar product \( \langle B | B \rangle \). We have
\begin{align}
\begin{split}
    \langle B | B \rangle &= \sum_{j=j_0}^{J-1} \sum_{\alpha=1}^{2j+1} (2j+1) = \sum_{j=j_0}^{J-1} (2j+1)^2 = \sum_{j=j_0}^{J-1} (4j^2 + 4j + 1) = 4 \sum_{j=j_0}^{J-1} j^2 + 4 \sum_{j=j_0}^{J-1} j + \sum_{j=j_0}^{J-1} 1 = \\
    &= 4 \frac{(J-1+2j_0)(J-j_0)(2J-1-2j_0)}{6} + 4 \frac{(J-j_0)(j_0+J-1)}{2} + J-j_0 \, .
\end{split}
\end{align}
It can easily be seen that in the limit \( J \to \infty \) we have
\begin{align}\label{mnEvhG7fghSuj6jiw1edn}
    \langle A | B \rangle = \dfrac{2\sqrt{2}}{\pi} J^{\frac{3}{2}} + \mathcal{O} \big( J^{\frac{1}{2}} \big)
\end{align}
and
\begin{align}\label{w35Sa65es98a75rdtfg8iuh876j}
    \langle B | B \rangle = \frac{4}{3} J^3 + \mathcal{O} \big( J^2 \big) \, .
\end{align}
It is interesting to see how much \( |A\rangle \) and \( |B\rangle \) overlap in the limit of large \(J\). From Eqs. \eqref{yfyghv7btuyh6uv7rbyghFFJn7}, \eqref{mnEvhG7fghSuj6jiw1edn} and \eqref{w35Sa65es98a75rdtfg8iuh876j} it follows that
\begin{align}
    \dfrac{\langle A | B \rangle}{\lVert A \rVert \lVert B \rVert} = \dfrac{\langle A | B \rangle}{\sqrt{\langle B | B \rangle}} = \dfrac{\sqrt{6}}{\pi} + \mathcal{O} \big( J^{-1} \big) \approx 0.78 + \mathcal{O} \big( J^{-1} \big) \, .
\end{align}

\section{Proof of Asymptotic Covariant Agreement of Multiple Observers}
\label{appendix:delta}
In this appendix, we provide a detailed proof of the asymptotic covariant agreement among multiple Bobs in the process of reference frame transmission. Our goal is to demonstrate that, in the limit of large \(J\), the reference frames estimated by different observers asymptotically covariantly agree with each other (covariant agreement should be understood in the sense described in Section \ref{sec:manyobservers}). In Appendix \ref{appendix:scalarproduct}, we have calculated  that the probability density of estimating \( g' \), given that the measured state is \( | A (g) \rangle \), is
\begin{align}\label{bftyghjb7iyugh67bGH}
    p(g' | g) = p(\theta) = \frac{2}{J-j_0+1} \frac{\sin^2 \left[ (J+j_0) \theta \right]}{\sin^2 \theta } \dfrac{\sin^2 \left( \dfrac{\pi}{J-j_0+1} \right) \cos^2 \left[ (J-j_0+1) \theta \right]}{\left[ \cos  2 \theta - \cos \left( \dfrac{\pi}{J-j_0+1} \right) \right]^2} \, ,
\end{align}
where \(\theta \in [0,\pi]\) is the hyperspherical angle, given by the following equation:
\begin{align}
    \cos\theta = \frac{1}{2} \Tr \left( g^{-1}g' \right) \, .
\end{align}
Let us note that
\begin{align}
    \Tr \left[ g^{-1}g' \right] = \Tr \left[ \left( g'^{-1} g \right)^{-1} \right] = \Tr \left[ g'^{-1} g \right] \, ,
\end{align}
where the last equality is a consequence of Eq. \eqref{rdbtfygj4576nhTJ32c0}. This implies that
\begin{align}\label{errtfhyg566gGRTDTFHxe}
    p(g' | g) = p(g | g') \, .
\end{align}
Let us recall that \( \mathcal{R}: \mathrm{SU(2)} \to \mathrm{SO(3)} \) is a map that assigns to an element of the \(\mathrm{SU(2)}\) group the corresponding element of the \( \mathrm{SO(3)} \) group. Let  \( f \in C^{\infty} \left[ \mathrm{SO(3)} \right] \) be a test function and \( g' \) be a fixed element of the \( \mathrm{SU(2)} \) group. We have
\begin{align}
    \int\displaylimits_{\mathrm{SU(2)}} \mathrm{d}g \, p( g' | g ) f \left[ \mathcal{R} (g) \right] = \int\displaylimits_{\mathrm{SU(2)}} \mathrm{d}g \, p( g^{-1} g'| \mathbb{1}_2 )  f \left[ \mathcal{R} (g) \right] = \int\displaylimits_{\mathrm{SU(2)}} \mathrm{d}h \, p( h | \mathbb{1}_2 ) f \left[ \mathcal{R} \left( g'h^{-1} \right) \right] = (\star) \, .
\end{align}
Let  \( (\theta,\psi,\varphi) \) be the hyperspherical coordinates corresponding to the group element \( h \in \mathrm{SU(2)} \). We define
\begin{align}\label{exrbftghj354678hhbgdGH6}
f_{g'}(\theta,\psi,\varphi) \equiv f \left[ \mathcal{R} \left( g'h^{-1} \right) \right] \, .
\end{align}
One can easily verify that
\begin{align}\label{ntygju657678nbTYYJ}
    f_{g'}(\pi - \theta, \pi - \psi, \varphi + \pi) = f \left[ \mathcal{R} \left( g'(-h)^{-1} \right) \right] = f \left[ \mathcal{R} \left( -g'h^{-1} \right) \right] = f \left[ \mathcal{R} \left( g'h^{-1} \right) \right] \, ,
\end{align}
where the first equality is a consequence of Eq. \eqref{rertgfhy467878GFuy} and the last equality is a consequence of Eq. \eqref{hytjyh5587678hjjYUJHG}. Equations \eqref{exrbftghj354678hhbgdGH6} and \eqref{ntygju657678nbTYYJ} imply that
\begin{align}\label{nyXuhj7PZyujertfUGIhg}
    f_{g'}(\theta,\psi,\varphi) = f_{g'}(\pi - \theta, \pi - \psi, \varphi + \pi) \, .
\end{align}
We have
\begin{align}
    (\star) = \frac{1}{2 \pi^2} \int_{0}^{\pi} \sin^2 \theta \, \mathrm{d}\theta \int_{0}^{\pi} \sin \psi \, \mathrm{d}\psi \int_{0}^{2 \pi} \mathrm{d}\varphi \, p(\theta) f_{g'}(\theta,\psi,\varphi) = \frac{2}{\pi} \int_{0}^{\pi} p(\theta) F_{g'}(\theta) \sin^2 \theta \, \mathrm{d}\theta
    = (\star \star) \, ,
\end{align}
where for \( \theta \in [0,\pi] \) we defined
\begin{align}\label{uhtdrfyhfgERGDTFH67fgrtGHT4HF}
    F_{g'}(\theta) \equiv \frac{1}{4\pi} \int_{0}^{\pi} \sin \psi \, \mathrm{d}\psi \int_{0}^{2 \pi} \mathrm{d}\varphi \, f_{g'}(\theta,\psi,\varphi) \, .
\end{align}
Let us note that
\begin{align}\label{rty88674hgBJHG9nhjA}
\begin{split}
    F_{g'}(\pi - \theta) &= \frac{1}{4\pi} \int_{0}^{\pi} \sin \psi \, \mathrm{d}\psi \int_{0}^{2 \pi} \mathrm{d}\varphi \, f_{g'}(\pi - \theta, \pi - \psi, \varphi + \pi) =\\
    &= \frac{1}{4\pi} \int_{0}^{\pi} \sin \psi'' \, \mathrm{d}\psi'' \int_{0}^{2 \pi} \mathrm{d}\varphi'' \, f_{g'}(\pi - \theta, \psi'', \varphi'') = F_{g'}(\theta) \, .
\end{split}
\end{align}
Continuing our main calculation, we get
\begin{align}\label{tfjhdghgjjhg7586D78}
\begin{split}
    (\star \star) &= \frac{2}{\pi} \int_{0}^{\frac{\pi}{2}} p(\theta) F_{g'}(\theta) \sin^2 \theta \, \mathrm{d}\theta + \frac{2}{\pi} \int_{\frac{\pi}{2}}^{\pi} p(\theta) F_{g'}(\theta) \sin^2 \theta \, \mathrm{d}\theta =\\
    &= \frac{2}{\pi} \int_{0}^{\frac{\pi}{2}} p(\theta) F_{g'}(\theta) \sin^2 \theta \, \mathrm{d}\theta + \frac{2}{\pi} \int_{\frac{\pi}{2}}^{\pi} p(\pi - \theta) F_{g'}(\pi - \theta) \sin^2 \left( \pi - \theta \right) \, \mathrm{d}\theta = (\circ) \, ,\\
\end{split}
\end{align}
where we used the identities \( p(\theta) = p(\pi - \theta) \) (a direct consequence of Eq. \eqref{bftyghjb7iyugh67bGH}) and \( F_{g'}(\theta) = F_{g'}(\pi - \theta) \) (see Eq. \eqref{rty88674hgBJHG9nhjA}). Then we substitute \( \theta'' = \pi - \theta \), change the variable of integration in the last integral, and obtain
\begin{align}
\begin{split}
    (\circ) = \frac{2}{\pi} \int\displaylimits_{0}^{\frac{\pi}{2}} p(\theta) F_{g'}(\theta) \sin^2 \theta \mathrm{d}\theta + \frac{2}{\pi} \int\displaylimits_{0}^{\frac{\pi}{2}} p(\theta'') F_{g'}(\theta'') \sin^2 \theta'' \mathrm{d}\theta'' = \frac{4}{\pi} \int\displaylimits_{0}^{\frac{\pi}{2}} p(\theta) F_{g'}(\theta) \sin^2 \theta \mathrm{d}\theta = (\circ \circ) \, .\\
\end{split}
\end{align}
Now we substitute the expression for \( p(\theta) \) (see Eq. \eqref{bftyghjb7iyugh67bGH}), which gives us
\begin{align}
    (\circ \circ) = \frac{8}{\pi} \frac{\sin^2 \left( \dfrac{\pi}{J-j_0+1} \right)}{J-j_0+1}  \int\displaylimits_{0}^{\frac{\pi}{2}} \dfrac{ \sin^2 \left[ (J+j_0) \theta \right] \cos^2 \left[ (J-j_0+1) \theta \right]}{\left[ \cos  2 \theta - \cos \left( \dfrac{\pi}{J-j_0+1} \right) \right]^2}  F_{g'}(\theta) \, \mathrm{d}\theta \, .
\end{align}
This in turn implies that
\begin{align}
    \lim_{J \to +\infty} \int\displaylimits_{\mathrm{SU(2)}} \mathrm{d}g \, p( g' | g ) f \left[ \mathcal{R} (g) \right] = \lim_{J \to +\infty} \frac{8\pi}{J^3} \int\displaylimits_{0}^{\frac{\pi}{2}} \dfrac{ \sin^2 \left[ (J+j_0) \theta \right] \cos^2 \left[ (J-j_0+1) \theta \right]}{\left[ \cos  2 \theta - \cos \left( \dfrac{\pi}{J-j_0+1} \right) \right]^2}  F_{g'}(\theta) \, \mathrm{d}\theta \, .
\end{align}
For large \(J\), the integrand can be simplified by noting that it is significantly different from zero only for \( \theta \approx 0 \). Therefore, we can expand it in \( \theta \), which leads to the following expression:
\begin{align}
    \lim_{J \to +\infty} \int\displaylimits_{\mathrm{SU(2)}} \mathrm{d}g \, p( g' | g ) f \left[ \mathcal{R} (g) \right] = \lim_{J \to +\infty} \frac{2\pi}{J^3} \int\displaylimits_{0}^{\frac{\pi}{2}} \dfrac{ \sin^2 \left( 2J\theta \right)}{\left( 2\theta^2 - \dfrac{\pi^2}{2J^2} \right)^2} F_{g'}(|\theta|) \, \mathrm{d}\theta = (\diamond)
\end{align}
The above integrand is symmetric, so we can integrate from $-\frac{\pi}{2}$ to $\frac{\pi}{2}$. We have
\begin{align}
    (\diamond) = \lim_{J \to +\infty} 8 \pi J \int\displaylimits_{0}^{\frac{\pi}{2}} \frac{\sin^2 \left( 2 J \theta \right)}{\left( 4 J^2 \theta^2 - \pi^2 \right)^2} F_{g'}(|\theta|) \, \mathrm{d}\theta = \lim_{J \to +\infty} 4 \pi J \int\displaylimits_{-\frac{\pi}{2}}^{\frac{\pi}{2}} \frac{\sin^2 \left( 2 J \theta \right)}{\left( 4 J^2 \theta^2 - \pi^2 \right)^2} F_{g'}(|\theta|) \, \mathrm{d}\theta = (\diamond \diamond) \, .
\end{align}
We define a new function
\begin{align}\label{twyknjhvnnoi546gtghhfFS}
    F_{g'}^{0}(\theta) =
    \begin{cases}
    F_{g'}(|\theta|) & \mathrm{for} \; \; \theta \in [-\frac{\pi}{2},\frac{\pi}{2}]\\
    0 & \mathrm{for} \; \; \theta \in \mathbb{R} \setminus [-\frac{\pi}{2},\frac{\pi}{2}]
    \end{cases} \, .
\end{align}
Then
\begin{align}
    (\diamond \diamond) &= \lim_{J \to +\infty} 4 \pi J \int\displaylimits_{-\infty}^{\infty} \frac{\sin^2 \left( 2 J \theta \right)}{\left( 4 J^2 \theta^2 - \pi^2 \right)^2} F_{g'}^{0}(\theta) \, \mathrm{d}\theta = \lim_{\varepsilon \to 0^+} \frac{2\pi}{\varepsilon} \int\displaylimits_{-\infty}^{+\infty} \frac{\sin^2 \left( \dfrac{\theta}{\varepsilon} \right)}{\left[ \left( \dfrac{\theta}{\varepsilon} \right)^2 - \pi^2 \right]^2} F_{g'}^{0}(\theta) \, \mathrm{d}\theta = (*) \, ,
\end{align}
where $\varepsilon = (2J)^{-1}$. We define
\begin{align}
    \eta_{\varepsilon}(\theta) = \frac{2\pi}{\varepsilon} \frac{\sin^2 \left( \dfrac{\theta}{\varepsilon} \right)}{\left[ \left( \dfrac{\theta}{\varepsilon} \right)^2 - \pi^2 \right]^2} \, ,
\end{align}
and we denote \( \eta(\theta) \equiv \eta_1(\theta) \). It is easy to see that
\begin{align}\label{model_delty_skalowanie}
    \eta_\varepsilon(\theta) = \varepsilon^{-1} \eta \left (\frac{\theta}{\varepsilon} \right) \, .
\end{align}
Moreover, we have
\begin{align}\label{model_delty_calka}
\begin{split}
    \int\displaylimits_{-\infty}^{+\infty} \eta(x) \, \mathrm{d}x &= 2\pi \int\displaylimits_{-\infty}^{+\infty} \frac{\sin^2 x}{\left( x^2 - \pi^2 \right)^2} \, \mathrm{d}x = \frac{1}{2 \pi} \int\displaylimits_{-\infty}^{+\infty} \left[ \frac{\sin^2 x}{\left( x - \pi \right)^2} + \frac{\sin^2 x}{\left( x + \pi \right)^2} - \frac{2\sin^2 x}{x^2 - \pi^2} \right] \, \mathrm{d}x =\\
    &= \frac{1}{\pi} \left( \int\displaylimits_{-\infty}^{+\infty} \frac{\sin^2 x}{x^2} \, \mathrm{d}x - \int\displaylimits_{-\infty}^{+\infty} \frac{\sin^2 x}{x^2 - \pi^2} \, \mathrm{d}x \right) = \frac{1}{\pi} \left( \pi - 0 \right) = 1 \, .\\
\end{split}
\end{align}
Equations \eqref{model_delty_skalowanie} and \eqref{model_delty_calka} show that
\begin{align}
    \lim_{\varepsilon\to 0^+} \eta_\varepsilon(\theta) = \delta (\theta) \, ,
\end{align}
where \( \delta \) is the Dirac delta function. This gives us
\begin{align}\label{gfhgjfvgdrt5655gg99zG}
    (*) = \int\displaylimits_{-\infty}^{+\infty} \delta (\theta) F_{g'}^{0}(\theta) \, \mathrm{d}\theta = F_{g'}^{0}(0) = F_{g'}(0) = \frac{1}{4\pi} \int_{0}^{\pi} \sin \psi \, \mathrm{d}\psi \int_{0}^{2 \pi} \mathrm{d}\varphi \, f_{g'}(0,\psi,\varphi) = (**) \, ,
\end{align}
where we have also substituted Eqs. \eqref{twyknjhvnnoi546gtghhfFS} and \eqref{uhtdrfyhfgERGDTFH67fgrtGHT4HF}. Regardless of the values of \( \psi \) and \( \varphi \), the hyperspherical angle \(\theta = 0\) corresponds to \( h = \mathbb{1}_2 \in \mathrm{SU(2)} \). Therefore, we can substitute Eq. \eqref{exrbftghj354678hhbgdGH6} into Eq. \eqref{gfhgjfvgdrt5655gg99zG}, and then we obtain
\begin{align}
        (**) = \frac{1}{4\pi} \int_{0}^{\pi} \sin \psi \, \mathrm{d}\psi \int_{0}^{2 \pi} \mathrm{d}\varphi \, f \left[ \mathcal{R} \left( g'\mathbb{1}_2^{-1} \right) \right] = f \left[ \mathcal{R} \left( g' \right) \right] \, ,
\end{align}
thus
\begin{align}\label{tdvryhgtfvhyugvfy5uyu6576yG}
    \lim_{J \to +\infty} \int\displaylimits_{\mathrm{SU(2)}} \mathrm{d}g \, p( g' | g ) f \left[ \mathcal{R} (g) \right] = f \left[ \mathcal{R} \left( g' \right) \right] \, .
\end{align}
The joint probability density that the \(i\)th Bob estimates \( g_i' \) is (cf. \eqref{fbhjhgjtj457678ghghdfhjgYTp7})
\begin{align}
    p(g_1', \dots, g_k') = \int\displaylimits_{\mathrm{SU(2)}} \mathrm{d}g \, p_0 \left[ \mathcal{R}(g) \right] \prod_{i=1}^{k} p( g_i' | g g_i ) \, .
\end{align}
Let \( f_i \in C^{\infty}\left[\mathrm{SO(3)}\right] \) be test functions. We have
\begin{align}\label{rnty567hy67ytgrdttyufgytuH}
    \idotsint\displaylimits_{\mathrm{SU(2)} \; \, \; \mathrm{SU(2)}} \mathrm{d}g'_1 \ldots \mathrm{d}g'_k \, p(g'_1, \ldots, g'_k) \prod_{i=1}^{k} f_i \left[ \mathcal{R} (g'_i) \right] = \int\displaylimits_{\mathrm{SU(2)}} \mathrm{d}g \, p_0 \left[ \mathcal{R}(g) \right] \prod_{i=1}^{k} \left[ \int\displaylimits_{\mathrm{SU(2)}} \mathrm{d}g'_i \, p( g_i' | g g_i ) f_i \left[ \mathcal{R} (g'_i) \right] \right] \, .
\end{align}
Equations \eqref{errtfhyg566gGRTDTFHxe} and \eqref{tdvryhgtfvhyugvfy5uyu6576yG} imply that
\begin{align}\label{ybrythju667uyughj8987}
    \lim_{J \to +\infty } \int\displaylimits_{\mathrm{SU(2)}} \mathrm{d}g'_i \, p( g_i' | g g_i ) f_i \left[ \mathcal{R} (g'_i) \right] =  \lim_{J \to +\infty } \int\displaylimits_{\mathrm{SU(2)}} \mathrm{d}g'_i \, p( g g_i | g_i' ) f_i \left[ \mathcal{R} (g'_i) \right] = f_i \left[ \mathcal{R} (g g_i) \right] \, .
\end{align}
Putting Eqs. \eqref{rnty567hy67ytgrdttyufgytuH} and \eqref{ybrythju667uyughj8987} together, we get
\begin{align}
    \lim_{J \to +\infty } \idotsint\displaylimits_{\mathrm{SU(2)} \; \, \; \mathrm{SU(2)}} \mathrm{d}g'_1 \ldots \mathrm{d}g'_k \, p(g'_1, \ldots, g'_k) \prod_{i=1}^{k} f_i \left[ \mathcal{R} (g'_i) \right] = \int\displaylimits_{\mathrm{SU(2)}} \mathrm{d}g \, p_0 \left[ \mathcal{R}(g) \right] \prod_{i=1}^{k} f_i \left[ \mathcal{R} (g g_i) \right] \, .
\end{align}
Now let
\begin{align}\label{btfjyghkyujh567uyhfghGFBFH7}
    f_{tot} \in C^{\infty} \left[ \bigtimes_{i=1}^{k} \mathrm{SO(3)} \right]
\end{align}
be a test function and
\begin{align}
    \mathcal{R}_k: \bigtimes_{i=1}^{k} \mathrm{SU(2)} \to \bigtimes_{i=1}^{k} \mathrm{SO(3)}: \left( g_1, \dots, g_k \right) \mapsto \left( \mathcal{R}(g_1), \dots, \mathcal{R}(g_k) \right) \, .
\end{align}
The space of test functions in the product form is a dense subspace of space of test functions in the form \eqref{btfjyghkyujh567uyhfghGFBFH7}, so
\begin{align}\label{tyjujtfyu67543676hGHHh}
\begin{split}
    \lim_{J \to +\infty }\idotsint\displaylimits_{\mathrm{SU(2)} \; \, \; \mathrm{SU(2)}} \mathrm{d}g'_1 \ldots \mathrm{d}g'_k \, p(g'_1, \ldots, g'_k) (f_{tot} \circ \mathcal{R}_k ) & \left( g'_1, \ldots, g'_k \right) =\\ = \int\displaylimits_{\mathrm{SU(2)}} \mathrm{d}g \, p_0 \left[ \mathcal{R}(g) \right] (f_{tot} \circ \mathcal{R}_k ) & \left( g g_1, \ldots, g g_k \right) \, .
\end{split}
\end{align}
We can rewrite Eq. \eqref{tyjujtfyu67543676hGHHh} in the following form (see also Eqs. \eqref{rrbty45677hygjbfsfF}, \eqref{bgh58768hjhnHNJJ78hyu4}, and \eqref{yvgjbh547687bcfgVGHJNH7d}):
\begin{align}\label{retygnjk6778Gyuhj6ASc}
\begin{split}
    \lim_{J \to +\infty }\idotsint\displaylimits_{\mathrm{\mathrm{SO(3)}} \; \, \; \mathrm{\mathrm{SO(3)}}} \mathrm{d}R'_1 \ldots \mathrm{d}R'_k \, p(R'_1, \ldots, R'_k) f_{tot} \left( R'_1, \ldots, R'_k \right)  = \int\displaylimits_{\mathrm{\mathrm{SO(3)}}} \mathrm{d}R \, p_0(R) f_{tot} \left( R R_1, \ldots, R R_k \right) \, .
\end{split}
\end{align}
This is the main result of this appendix. Expressed in terms of Dirac delta functions, it reads:
\begin{align}\label{appagreement}
    \lim_{J \to +\infty } p(R'_1, \ldots, R'_k) = \int\displaylimits_{\mathrm{\mathrm{SO(3)}}} \mathrm{d}R \, p_0(R) \prod_{i=1}^{k} \delta (R R_i, R_i') \, .
\end{align}
A detailed physical interpretation of this result can be found in Section \ref{sec:agreement}.

\section{Derivation of the Post-Measurement State and Trace Distance Bounds}
\label{appendix:disturbance}
In this appendix, we derive the explicit form of the quantum state resulting from a measurement process and analyze its asymptotic behavior in the limit of large total angular momentum \(J\). Our goal is twofold: first, to obtain a rigorous expression for the post-measurement state; and second, to quantify how closely it approximates the original (pre-measurement) state by evaluating the trace distance between them. The state that encodes and objectifies Alice's reference frame prior to measurement is given by the following expression (cf. \eqref{brtfgh7uyg687hjhTGhj7}):
\begin{align}\label{appendixbrtfgh7uyg687hjhTGhj7}
    \rho = \int\displaylimits_{\mathrm{SU(2)}} \mathrm{d}g \, p_0 &\left[ \mathcal{R}(g) \right] | g \rangle \langle g | \otimes \bigotimes_{i=1}^{k} | A(g g_i) \rangle \langle A(g g_i) | \, .
\end{align}
The composite measurement is given by the following POVM (cf. \eqref{compositePOVM}):
\begin{align}\label{appendixcompositePOVM}
    M(g'_1, \ldots, g'_k) = \openone_R \otimes \bigotimes_{i=1}^{k} | B(g'_i) \rangle \langle B(g'_i) | \, , 
\end{align}
satisfying the following completeness relation:
\begin{align}
    \idotsint\displaylimits_{\mathrm{SU(2)} \; \, \; \mathrm{SU(2)}} \mathrm{d}g'_1 \ldots \mathrm{d}g'_k \, M(g'_1, \ldots, g'_k) = \mathbb{1} \, .
\end{align}
The joint probability density of estimating \( (g'_1, \ldots, g'_k) \) is given by the Born rule and equals
\begin{align}\label{appendixfbhjhgjtj457678ghghdfhjgYTp7}
    p(g_1', \dots, g_k') = \mathrm{Tr} \left[ M(g'_1, \ldots, g'_k) \rho \right] = \int\displaylimits_{\mathrm{SU(2)}} \mathrm{d}g \, p_0 \left[ \mathcal{R}(g) \right] \mathrm{Tr} \left( |g \rangle \langle g| \right)  \prod_{i=1}^{k} p( g_i' | g g_i ) \, ,
\end{align}
where
\begin{align}\label{appendixprobdensity}
    p( g_i' | g g_i ) = |\langle A(g g_i) | B(g_i') \rangle |^2 \, .
\end{align}
The measurement operators are the following:
\begin{align}\label{measopappendix}
    \sqrt{M(g'_1, \ldots, g'_k)} = \openone_R \otimes \bigotimes_{i=1}^{k} \sqrt{| B(g'_i) \rangle \langle B(g'_i) |} = \openone_R \otimes \bigotimes_{i=1}^{k} \frac{| B(g'_i) \rangle \langle B(g'_i) |}{\| B \|} \, .
\end{align}
Strictly speaking, when taking the square root of the effects \(M(g'_1, \ldots, g'_k) \), there is a freedom up to a unitary rotation. The original proposal \cite{bagan, Chiribella_2004} did not fix this, as it was not needed. Here, we interpret the decoding algorithm as consisting simply of projections onto the \(B\)-vectors, without any additional unitary rotations (which would have to satisfy a hierarchy of conditions, stemming from partial tracing of \(M(g'_1, \ldots, g'_k)\)). With this assumption, the conditional post-measurement state is:
\begin{align}
    \rho' (g'_1, \ldots, g'_k) = \frac{\sqrt{M(g'_1, \ldots, g'_k)} \rho \sqrt{M(g'_1, \ldots, g'_k)}}{\Tr \left[ \sqrt{M(g'_1, \ldots, g'_k)} \rho \sqrt{M(g'_1, \ldots, g'_k)} \right]} \, .
\end{align}
The average post-measurement state is given by the following expression:
\begin{align}\label{averagepost1}
\begin{split}
    \rho' &= \idotsint\displaylimits_{\mathrm{SU(2)} \; \, \; \mathrm{SU(2)}} \mathrm{d}g'_1 \ldots \mathrm{d}g'_k \, p(g_1', \dots, g_k') \rho' (g'_1, \ldots, g'_k) =\\
    &= \idotsint\displaylimits_{\mathrm{SU(2)} \; \, \; \mathrm{SU(2)}} \mathrm{d}g'_1 \ldots \mathrm{d}g'_k \, \mathrm{Tr} \left[ M(g'_1, \ldots, g'_k) \rho \right] \frac{\sqrt{M(g'_1, \ldots, g'_k)} \rho \sqrt{M(g'_1, \ldots, g'_k)}}{\Tr \left[ \sqrt{M(g'_1, \ldots, g'_k)} \rho \sqrt{M(g'_1, \ldots, g'_k)} \right]} = (*) \, .
\end{split}
\end{align}
The traces in the numerator and denominator cancel out, yielding
\begin{align}
\begin{split}
    (*) &= \idotsint\displaylimits_{\mathrm{SU(2)} \; \, \; \mathrm{SU(2)}} \mathrm{d}g'_1 \ldots \mathrm{d}g'_k \, \sqrt{M(g'_1, \ldots, g'_k)} \rho \sqrt{M(g'_1, \ldots, g'_k)} =\\
    &= \idotsint\displaylimits_{\mathrm{SU(2)} \; \, \; \mathrm{SU(2)}} \mathrm{d}g'_1 \ldots \mathrm{d}g'_k \, \left( \openone_R \otimes \bigotimes_{i=1}^{k} \frac{| B(g'_i) \rangle \langle B(g'_i) |}{\| B \|} \right)  \rho \left( \openone_R \otimes \bigotimes_{i=1}^{k} \frac{| B(g'_i) \rangle \langle B(g'_i) |}{\| B \|} \right) = (**) \, ,
\end{split}
\end{align}
where we have also substituted the expression for \( \sqrt{M(g'_1, \ldots, g'_k)} \) (Eq. \eqref{measopappendix}). Finally, we substitute the expression for \( \rho \) (Eq. \eqref{appendixbrtfgh7uyg687hjhTGhj7}), identify the modulus squared of the scalar product with the conditional probability (Eq. \eqref{appendixprobdensity}), and obtain
\begin{align}\label{postmstateapp}
\begin{split}
    (**) &= \int\displaylimits_{\; \, \; \mathrm{SU(2)}} \mathrm{d}g \, p_0 \left[ \mathcal{R}(g) \right] | g \rangle \langle g | \otimes \idotsint\displaylimits_{\mathrm{SU(2)} \; \, \; \mathrm{SU(2)}} \mathrm{d}g'_1 \ldots \mathrm{d}g'_k \, \bigotimes_{i=1}^{k} \frac{| B(g'_i) \rangle |\langle A(g g_i) | B(g_i') \rangle |^2 \langle B(g'_i) |}{\| B \|^2} =\\
    &= \int\displaylimits_{\; \, \; \mathrm{SU(2)}} \mathrm{d}g \, p_0 \left[ \mathcal{R}(g) \right] | g \rangle \langle g | \otimes \bigotimes_{i=1}^{k} \left( \, \int\displaylimits_{\; \, \; \mathrm{SU(2)}} \mathrm{d}g'_i \, p( g_i' | g g_i ) \frac{| B(g'_i) \rangle \langle B(g'_i) |}{\| B \|^2} \right) \, .
\end{split}
\end{align}
Combining Eqs. \eqref{appendixbrtfgh7uyg687hjhTGhj7} and \eqref{postmstateapp}, we obtain the difference between the post-measurement and pre-measurement states:
\begin{align}\label{nyuIMHdrgf576h323sa1}
    \rho - \rho' = \int\displaylimits_{\mathrm{SU(2)}} \mathrm{d}g \, p_0 \left[ \mathcal{R}(g) \right] | g \rangle \langle g | \, \otimes C(g) \, ,
\end{align}
where
\begin{align}
    C(g) \equiv \bigotimes_{i=1}^{k} | A(g g_i) \rangle \langle A(g g_i) | - \bigotimes_{i=1}^{k} \left( \, \int\displaylimits_{\; \, \; \mathrm{SU(2)}} \mathrm{d}g'_i \, p( g_i' | g g_i ) \frac{| B(g'_i) \rangle \langle B(g'_i) |}{\| B \|^2} \right) \, .
\end{align}
We want to estimate the trace distance between \( \rho \) and \( \rho' \), defined as:
\begin{align}\label{tjhy646789923239hghYP}
   T(\rho,\rho') \equiv \frac{1}{2} \|\rho - \rho'\|_{1} = \frac{1}{2} \Tr \left[ \sqrt{(\rho-\rho')^\dagger (\rho-\rho')} \right] \, .
\end{align}
First, we substitute Eq. \eqref{nyuIMHdrgf576h323sa1} into Definition \eqref{tjhy646789923239hghYP} and obtain the following expression for the trace norm of \( \rho - \rho' \):
\begin{align}
    \left\| \rho - \rho' \right\|_{1} = \Tr \left[ \int\displaylimits_{\; \, \; \mathrm{SU(2)}} \mathrm{d}g \, \int\displaylimits_{\; \, \; \mathrm{SU(2)}} \mathrm{d}h \, p_0 \left[ \mathcal{R}(g) \right] p_0 \left[ \mathcal{R}(h) \right] | g \rangle \langle g | h \rangle \langle h | \otimes  C^\dagger(g)  C(h) \right]^{\frac{1}{2}} = ( \circ ) \, .
\end{align}
We now use the identity \( \langle g | h \rangle = \delta(g,h) \) (cf. \eqref{diracdeltanormalization}) and get
\begin{align}\label{ui8766hghbdtfgjhNGF}
    ( \circ ) &= \Tr \left[ \int\displaylimits_{\; \, \; \mathrm{SU(2)}} \mathrm{d}g \, \Big( p_0 \left[ \mathcal{R}(g) \right] \Big)^2 | g \rangle \langle g | \otimes  C^\dagger(g)  C(g) \right]^{\frac{1}{2}} = ( \circ \circ ) \, .
\end{align}
For each g, we can spectrally decompose the operator \( C^\dagger(g)  C(g) \) as
\begin{align}\label{465677hg099gf62s}
    C^\dagger(g)  C(g) = \sum_{\alpha} \lambda_{\alpha}(g) | v_{\alpha}(g) \rangle \langle v_{\alpha}(g)| \, .
\end{align}
Substituting Eq. \eqref{465677hg099gf62s} into Eq. \eqref{ui8766hghbdtfgjhNGF}, we obtain
\begin{align}
\begin{split}
    ( \circ \circ ) &= \Tr \left[ \int\displaylimits_{\; \, \; \mathrm{SU(2)}} \mathrm{d}g \, \Big( p_0 \left[ \mathcal{R}(g) \right] \Big)^2 | g \rangle \langle g | \otimes \sum_{\alpha} \lambda_{\alpha}(g) | v_{\alpha}(g) \rangle \langle v_{\alpha}(g)| \right]^{\frac{1}{2}} =\\
    &= \Tr \left[ \sum_{\alpha} \int\displaylimits_{\; \, \; \mathrm{SU(2)}} \mathrm{d}g \, \Big( p_0 \left[ \mathcal{R}(g) \right] \Big)^2 \lambda_{\alpha}(g) | g \rangle \langle g | \otimes | v_{\alpha}(g) \rangle \langle v_{\alpha}(g)| \right]^{\frac{1}{2}} = ( \bullet ) \, .
\end{split}
\end{align}
We compute the square root of the operator and obtain the following expression:
\begin{align}
\begin{split}
    ( \bullet ) &= \Tr \left[ \, \sum_{\alpha} \int\displaylimits_{\; \, \; \mathrm{SU(2)}} \mathrm{d}g \, p_0 \left[ \mathcal{R}(g) \right] \sqrt{\lambda_{\alpha}(g)} | g \rangle \langle g | \otimes | v_{\alpha}(g) \rangle \langle v_{\alpha}(g)| \right] =\\
    &= \int\displaylimits_{\mathrm{SU(2)}} \mathrm{d}g \, p_0 \left[ \mathcal{R}(g) \right] \Tr \left[ | g \rangle \langle g | \otimes \sum_{\alpha} \sqrt{\lambda_{\alpha}(g)} | v_{\alpha}(g) \rangle \langle v_{\alpha}(g)| \right] = ( \bullet \bullet ) \, .
\end{split}
\end{align}
Finally, we express the trace of the tensor product as a product of traces and recognize the sum as the trace norm of \(C(g)\), resulting in the following expression:
\begin{align}\label{apptracenorm}
    ( \bullet \bullet) = \int\displaylimits_{\mathrm{SU(2)}} \mathrm{d}g \, p_0 \left[ \mathcal{R}(g) \right] \Tr \left( |g \rangle \langle g| \right) \sum_{\alpha} \sqrt{\lambda_{\alpha}(g)} = \int\displaylimits_{\mathrm{SU(2)}} \mathrm{d}g \, p_0 \left[ \mathcal{R}(g) \right] \Tr \left( |g \rangle \langle g| \right) \left\| C(g) \right\|_{1} \, .
\end{align}
Similarly to the case of calculating the joint probability \( p(g_1', \dots, g_k') \) (see Eqs. \eqref{btngyjjrth5467yubhHGdfgh6} and \eqref{fbhjhgjtj457678ghghdfhjgYTp7}), we omit \( \Tr \left( |g \rangle \langle g| \right) \) and redefine the trace norm of \( \rho - \rho' \) as follows:
\begin{align}\label{redefinedtracenorm}
    \left\| \rho - \rho' \right\|_{1} = \int\displaylimits_{\mathrm{SU(2)}} \mathrm{d}g \, p_0 \left[ \mathcal{R}(g) \right] \left\| C(g) \right\|_{1} = \int\displaylimits_{\mathrm{SU(2)}} \mathrm{d}g \, p_0 \left[ \mathcal{R}(g) \right] \Big\| | \psi \rangle \langle \psi | - \sigma \Big\|_{1} \, ,
\end{align}
where
\begin{align}\label{defpsi}
    | \psi \rangle \equiv \bigotimes_{i=1}^{k} | A(g g_i) \rangle
\end{align}
and
\begin{align}\label{defsigma}
    \sigma \equiv \bigotimes_{i=1}^{k} \left( \, \int\displaylimits_{\; \, \; \mathrm{SU(2)}} \mathrm{d}g'_i \, p( g_i' | g g_i ) \frac{| B(g'_i) \rangle \langle B(g'_i) |}{\| B \|^2} \right) \, .
\end{align}
It is easy to see that
\begin{align}
    \Tr \sigma &= \prod_{i=1}^k \Tr \left[ \int\displaylimits_{\; \, \; \mathrm{SU(2)}} \mathrm{d}g'_i \, p( g_i' | g g_i ) \frac{| B(g'_i) \rangle \langle B(g'_i) |}{\| B \|^2} \right] =\\
    &= \prod_{i=1}^k \, \int\displaylimits_{\; \, \; \mathrm{SU(2)}} \mathrm{d}g'_i \, p( g_i' | g g_i ) \Tr \left[ \frac{| B(g'_i) \rangle \langle B(g'_i) |}{\| B \|^2} \right] = \prod_{i=1}^k \, \int\displaylimits_{\; \, \; \mathrm{SU(2)}} \mathrm{d}g'_i \, p( g_i' | g g_i ) = \prod_{i=1}^k 1 = 1 \, ,
\end{align}
i.e. \( \sigma \) is a properly normalized density operator. In our further calculations, we aim to estimate \( \left\| | \psi \rangle \langle \psi | - \sigma \right\|_{1} \) using the following inequalities between the trace distance and the fidelity (known as the Fuchs–van de Graaf inequalities, see \cite{fuchs} and \cite{Nielsen_Chuang_2010}):
\begin{align}\label{fidelitytracedistanceinequality}
     1 - F \left( |\psi \rangle \langle \psi | , \sigma \right) \leq \frac{1}{2} \Big\| | \psi \rangle \langle \psi | - \sigma \Big\|_{1} \leq \sqrt{1 - F \left( |\psi \rangle \langle \psi | , \sigma \right)} \, .
\end{align}
The lower bound is tighter than \( 1 - \sqrt{F \left( |\psi \rangle \langle \psi | , \sigma \right)} \) in \cite{fuchs}, because one of our states is pure \cite{Nielsen_Chuang_2010}. Let us recall that for arbitrary density operators \(\rho_1\) and \(\rho_2\), their fidelity is defined as
\begin{align}
    F(\rho_1, \rho_2) \equiv \left( \Tr \sqrt{\sqrt{\rho_1} \rho_2 \sqrt{\rho_1}} \right)^2 \, .
\end{align}
In our case, \( \rho_1 = |\psi \rangle \langle \psi | \) and \( \rho_2 = \sigma \). The fidelity then simplifies to
\begin{align}
    F \left( |\psi \rangle \langle \psi |, \sigma \right) = \langle \psi | \sigma |\psi \rangle \, .
\end{align}
We substitute the definitions of \( | \psi \rangle \) (Eq. \eqref{defpsi}) and \( \sigma \) (Eq. \eqref{defsigma}) into the above expression and obtain
\begin{align}\label{fidelityproduct}
\begin{split}
    F \left( |\psi \rangle \langle \psi | , \sigma \right) &= \left( \bigotimes_{i=1}^{k} \langle A(g g_i) | \right) \left( \bigotimes_{i=1}^{k}  \int\displaylimits_{\; \, \; \mathrm{SU(2)}} \mathrm{d}g'_i \, p( g_i' | g g_i ) \frac{| B(g'_i) \rangle \langle B(g'_i) |}{\| B \|^2} \right) \left( \bigotimes_{i=1}^{k} | A(g g_i) \rangle \right) =\\
    &= \prod_{i=1}^{k} \left( \, \int\displaylimits_{\; \, \; \mathrm{SU(2)}} \mathrm{d}g'_i \, p( g_i' | g g_i ) \frac{|\langle A(g g_i) | B(g'_i) \rangle|^2}{\| B \|^2} \right) = \prod_{i=1}^{k}  \left( \, \int\displaylimits_{\; \, \; \mathrm{SU(2)}} \mathrm{d}g'_i \, \frac{p( g_i' | g g_i )^2}{\| B \|^2} \right) \, .
\end{split}
\end{align}
Our goal now is to evaluate the above integrals. We have
\begin{align}
    \int\displaylimits_{\mathrm{SU(2)}} \mathrm{d}g'_i \, \frac{p( g_i' | g g_i )^2}{\| B \|^2} = \int\displaylimits_{\mathrm{SU(2)}} \mathrm{d}g'_i \, \frac{p( g_i^{-1} g^{-1} g_i' | \mathbb{1}_2 )^2}{\| B \|^2} = \int\displaylimits_{\mathrm{SU(2)}} \mathrm{d}h \, \frac{p( h | \mathbb{1}_2 )^2}{\| B \|^2} = ( \star ) \, .
\end{align}
Let  \( (\theta,\psi,\varphi) \) be the hyperspherical coordinates corresponding to the group element \( h \in \mathrm{SU(2)} \). Using Formula \eqref{njcxdjncdxjnhcdxnhjiuwf}, we obtain:
\begin{align}
\begin{split}
    (\star) &= \frac{1}{2 \pi^2} \int_{0}^{\pi} \sin^2 \theta \, \mathrm{d}\theta \int_{0}^{\pi} \sin \psi \, \mathrm{d}\psi \int_{0}^{2 \pi} \mathrm{d}\varphi \, \frac{p( \theta )^2}{\| B \|^2} = \frac{2}{\pi} \frac{1}{\| B \|^2} \int_{0}^{\pi} p(\theta)^2 \sin^2 \theta \, \mathrm{d}\theta =\\
    &= \frac{4}{\pi} \frac{1}{\| B \|^2} \int_{0}^{\frac{\pi}{2}} p(\theta)^2 \sin^2 \theta \, \mathrm{d}\theta = ( \star \star ) \, ,
\end{split}
\end{align}
where we used the identity \( p(\theta) = p(\pi - \theta) \) (cf. \eqref{tfjhdghgjjhg7586D78}). Subsequently, we substitute the expression for \( p(\theta) \) (Eq. \eqref{bftyghjb7iyugh67bGH}) and get
\begin{align}
    ( \star \star) = \frac{16}{\pi} \frac{1}{\| B \|^2} \frac{\sin^4 \left( \dfrac{\pi}{J-j_0+1} \right)}{(J-j_0+1)^2}  \int\displaylimits_{0}^{\frac{\pi}{2}} \dfrac{ \sin^4 \left[ (J+j_0) \theta \right] \cos^4 \left[ (J-j_0+1) \theta \right]}{ \sin^2 \theta \left[ \cos  2 \theta - \cos \left( \dfrac{\pi}{J-j_0+1} \right) \right]^4} \, \mathrm{d}\theta \, .
\end{align}
This implies that
\begin{align}
    \lim_{J \to +\infty} \int\displaylimits_{\mathrm{SU(2)}} \mathrm{d}g'_i \, \frac{p( g_i' | g g_i )^2}{\| B \|^2} = \lim_{J \to +\infty} \frac{12\pi^3}{J^9} \int\displaylimits_{0}^{\frac{\pi}{2}} \dfrac{ \sin^4 \left[ (J+j_0) \theta \right] \cos^4 \left[ (J-j_0+1) \theta \right]}{ \sin^2 \theta \left[ \cos  2 \theta - \cos \left( \dfrac{\pi}{J-j_0+1} \right) \right]^4} \, \mathrm{d}\theta \, ,
\end{align}
where we used the fact that asymptotically \( \| B \|^2 = \frac{4}{3} J^3 \) (cf. \eqref{w35Sa65es98a75rdtfg8iuh876j}). For large \(J\), the integrand can be simplified by noting that it is significantly different from zero only for \( \theta \approx 0 \). Therefore, we can expand it in \( \theta \), which leads to the following expression:
\begin{align}
    \lim_{J \to +\infty} \int\displaylimits_{\mathrm{SU(2)}} \mathrm{d}g'_i \, \frac{p( g_i' | g g_i )^2}{\| B \|^2} = \lim_{J \to +\infty} \frac{3 \pi^3}{4 J^9} \int\displaylimits_{0}^{\frac{\pi}{2}} \dfrac{ \sin^4 \left( 2J\theta \right)}{ \theta^2 \left( 2\theta^2 - \dfrac{\pi^2}{2J^2} \right)^4} \, \mathrm{d}\theta = (\diamond) \, .
\end{align}
Since the integrand is an even function of \( \theta \), we may extend the limits of integration symmetrically to \( \theta \in \left[ -\frac{\pi}{2}, \frac{\pi}{2} \right] \). Thus
\begin{align}\label{lambdafinal}
\begin{split}
    (\diamond) &= \lim_{J \to +\infty} \frac{3 \pi^3}{8 J^9} \int\displaylimits_{-\frac{\pi}{2}}^{\frac{\pi}{2}} \dfrac{ \sin^4 \left( 2J\theta \right)}{ \theta^2 \left( 2\theta^2 - \dfrac{\pi^2}{2J^2} \right)^4} \, \mathrm{d}\theta = \lim_{J \to +\infty} 12 \pi^3 \int\displaylimits_{-\frac{\pi}{2}}^{\frac{\pi}{2}} \dfrac{ \sin^4 \left( 2J\theta \right)}{ \left( 2 J \theta \right)^2 \left[ \left( 2 J \theta \right)^2 - \pi^2 \right]^4} \, 2J \mathrm{d} \theta =\\
    &= \lim_{J \to +\infty} 12 \pi^3 \int\displaylimits_{-J\pi}^{J\pi} \dfrac{ \sin^4 x}{ x^2 \left( x^2 - \pi^2 \right)^4} \, \mathrm{d} x = 12 \pi^3 \int\displaylimits_{-\infty}^{+\infty} \dfrac{ \sin^4 x}{ x^2 \left( x^2 - \pi^2 \right)^4} \, \mathrm{d} x \equiv \lambda \approx 0.236 \, .
\end{split}
\end{align}
The value of the last integral above has been evaluated and is denoted by \( \lambda \). Equations \eqref{fidelityproduct} and \eqref{lambdafinal} imply that
\begin{align}\label{fidelitylimit}
    \lim_{J \to +\infty} F \left( |\psi \rangle \langle \psi | , \sigma \right) = \lim_{J \to +\infty} \, \prod_{i=1}^{k}  \left( \, \int\displaylimits_{\; \, \; \mathrm{SU(2)}} \mathrm{d}g'_i \, \frac{p( g_i' | g g_i )^2}{\| B \|^2} \right) = \lambda^k \, .
\end{align}
It turns out that, in the limit \( J \to \infty \), the trace distance is bounded from below and above by the following expressions, as implied by Eqs. \eqref{fidelitytracedistanceinequality} and \eqref{fidelitylimit}:
\begin{align}\label{tracedistanceestimated1}
     1 - \lambda^k \leq \lim_{J \to +\infty} \frac{1}{2} \Big\| | \psi \rangle \langle \psi | - \sigma \Big\|_{1} \leq \sqrt{1 - \lambda^k} \, .
\end{align}
The result above, combined with Eq. \eqref{redefinedtracenorm}, gives us the following estimate of the trace distance between the post-measurement state and the original (pre-measurement) state:
\begin{align}\label{tracedistanceestimated2}
     1 - \lambda^k \leq \lim_{J \to +\infty} \frac{1}{2} \left\| \rho - \rho' \right\|_{1} \leq \sqrt{1 - \lambda^k} \, .
\end{align}
This shows that there is already a significant asymptotic disturbance for a single Bob (\(k=1\)), and it increases rapidly with the number of Bobs (the trace distance tends to \(1\) as \(k\) tends to infinity).

The calculations presented in this appendix allow us to straightforwardly extend the original reference frame transmission protocol to include an analysis of the disturbance to the state -- a consideration that was not addressed in \cite{Chiribella_2004}. The pre-measurement state (with respect to Bob’s orthogonal trihedron) is the following (see Eq. \eqref{02jdn3n3idn3di9m2sz}):
\begin{align}\label{premeasapp0}
    \rho_0 = | A(g_0) \rangle \langle A(g_0) | \, .
\end{align}
The measurement operators take the form (see Eqs. \eqref{mdj7suhe56wyegds2} and \eqref{mx39xnj39v548z20fafU}):
\begin{align}
    \sqrt{M(g')} = \sqrt{| B(g') \rangle \langle B(g') |} = \frac{| B(g') \rangle \langle B(g') |}{\| B \|} \, .
\end{align}
The average post-measurement state is calculated in the same manner as in the case with multiple observers (see Eqs. \eqref{averagepost1}--\eqref{postmstateapp}):
\begin{align}\label{postmeasapp0}
    \rho'_0 = \int\displaylimits_{\mathrm{SU(2)}} \mathrm{d}g' \, \sqrt{M(g')} \rho_0 \sqrt{M(g')} = \int\displaylimits_{\mathrm{SU(2)}} \mathrm{d}g' \, p(g' | g_0) \frac{| B(g') \rangle \langle B(g') |}{\| B \|^2} \, .
\end{align}
Comparing Eq. \eqref{premeasapp0} with Eq. \eqref{defpsi} and Eq. \eqref{postmeasapp0} with Eq. \eqref{defsigma}, we see that, in the limit \( J \to \infty \), the trace distance between the post-measurement state and the pre-measurement state can be estimated as follows (see Eq. \eqref{tracedistanceestimated1}):
\begin{align}\label{tracedistanceestimated0}
     1 - \lambda \leq \lim_{J \to +\infty} \frac{1}{2} \left\| \rho_0 - \rho_0' \right\|_{1} \leq \sqrt{1 - \lambda} \, .
\end{align}

\end{document}